%% file: d_d_arxiv.tex
\documentclass[aps,prx, superscriptaddress,amsmath,amssymb,floatfix, notitlepage, bm,braket, singlecolumn, preprint, longbibliography]
{revtex4-1}
\pdfoutput=1
\usepackage{times}
\usepackage{graphics}
\usepackage{graphicx}
\usepackage{color}
\usepackage{bm}
\usepackage{braket}
\usepackage{mathtools}
\usepackage{mathcomp}
\usepackage[caption=false,justification=raggedright,position=top,singlelinecheck=off]{subfig}
\usepackage{soul}
\usepackage{tabularx}
\usepackage{calrsfs}
\usepackage[vcentermath]{youngtab}
\usepackage[mathscr]{euscript}
\usepackage[cp1252]{inputenc}
\usepackage[normalem]{ulem}
\usepackage[dvipsnames]{xcolor}

\usepackage{hyperref}
\hypersetup{
colorlinks=true,
        linkcolor=blue,
        citecolor=blue,
        urlcolor=blue,
}

\usepackage{upgreek}

\newcommand{\dn}{\downarrow}

\newcommand{\up}{\uparrow}

\newcommand{\enni}{\noindent}
\newcommand{\enbe}{\begin{equation}}
\newcommand{\enee}{\end{equation}}
\newcommand{\enba}{\begin{align}}
\newcommand{\enea}{\end{align}}

\newcommand{\enhe}{$^{3}$He~}
\newcommand{\enst}{$s\tau_{3}$~}
\newcommand{\enht}{\bm{\hat{k}}}

\begin{document}

\title{
Multiorbital singlet pairing and 
$d+d$ superconductivity
}
\author{Emilian M.\ Nica}
\email{enica@asu.edu}
\affiliation{Department of Physics,
Arizona State University, 
Box 871504, Tempe 85287-1504, AZ, USA}
\author{Qimiao Si}
\email{qmsi@rice.edu}
\affiliation{Department of Physics and Astronomy, Rice University, 6100 Main St, Houston 77005 TX, USA}
\affiliation{Rice Center for Quantum Materials, Rice University, 6100 Main St, Houston 77005 TX, USA}
\date{\today}

\begin{abstract}
Recent experiments in multiband Fe-based and heavy-fermion superconductors have challenged the long-held dichotomy 
between simple $s$- and $d$-wave spin-singlet pairing states.
 Here, we advance several time-reversal-invariant irreducible pairings 
that go beyond the standard singlet functions through a matrix structure in the band/orbital space,
and elucidate their naturalness in multiband systems. We consider the $s\tau_{3}$ multiorbital superconducting state
for Fe-chalcogenide superconductors. This state, corresponding to  a $d+d$ intra- and inter-band pairing, is shown to
contrast with the more familiar $d +\text{i}d$ state in a way analogous to how the B- triplet pairing phase of \enhe superfluid
differs from its A- phase counterpart. In addition, we construct an analogue of the $s\tau_{3}$ pairing for the 
heavy-fermion  superconductor CeCu$_{2}$Si$_{2}$, using degrees-of-freedom that incorporate spin-orbit coupling.
Our results lead to the proposition that $d$-wave superconductors in correlated multiband systems
will generically have a fully-gapped Fermi surface
 when they are examined at sufficiently low energies.
\end{abstract}

\maketitle

\section*{Introduction}
\label{sec:intro}

Unconventional superconductivity of strongly correlated systems
is centrally important in condensed matter physics, with the
symmetry of the superconducting order parameter 
being a key issue of the field. This question appears to have reached a consensus in some notable instances. 
An example is the $d$-wave symmetry for the Cooper pairs
 in the well-studied Cu-based 
 superconductors (SCs) \cite{Lee_2006,Scalapino_2012}.
However, the pairing symmetry remains enigmatic in other classes of strongly correlated materials. 
For singlet superconductivity, the long-held dichotomy is between 
fully-gapped $s$- and 
nodal $d$-wave pairing states.
However, 
it has been increasingly recognized that multi- band/orbital systems are
inherently richer 
 for pairings \cite{Agterberg_2017, Ramires_2018}.
A canonical setting 
for multiorbital spin-singlet pairings
is the Fe-based 
 superconductors \cite{Hosono,Johnston2011,Dai_RMP2015,Yi2017,Si2016,Wang_Lee2011,Hirschfeld2016},
 especially for the Fe-chalcogenide cases.
 Here,
 the discovery of an orbital-selective Mott crossover in the normal state \cite{Yi_prl2013,Yu_prl2013}
motivated the notion of 
 orbital-selective superconductivity \cite{Yu_Zhu_Si}.
The latter opens up the possibilities for a variety of
orbital-dependent pairing states,
which have been studied in recent years
both theoretically \cite{Yin_Haule_Kotliar, Coleman, Nica_Yu, Kreisel2017,HY_Hu2019, Hu_PRR_2020}
and experimentally \cite{C_Zhang2013, Sprau, Shibauchi_arxiv_2020}.
In addition,
heavy fermion SCs, a class that includes about fifty members, have emerged 
as another prominent setting for singlet pairing states beyond the 
traditional possiblities
\cite{Smidman}.

 Recent experiments have 
 directly
 challenged the
conventional $s$- and $d$-wave dichotomy. 
 In alkaline Fe-selenides, 
 inelastic neutron scattering~\cite{Park, Friemel} revealed signatures of in-gap spin resonances, 
 whose characteristic wavevectors qualify them as
  a typical indicator of  sign-changing $d$-wave order parameters~\cite{Eschrig, Stockert_Nat_Phys_2011,Maier_PRB_2011, Dai_RMP2015,Si2016}. 
  By contrast, ARPES studies have indicated fully-gapped 
  superconductivity~\cite{Mou, Wang_2011, Xu, Wang_2012},
 even for a Fermi pocket near the center of the two-dimensional Brillouin zone 
 (BZ), which appears consistent with $s$- wave symmetry. Understanding the Fe-chalcogenide SC is crucially important, since the
 Fe-based superconductivity
 with the highest superconducting transition temperature ($T_{\text{c}}$) occurs in this category.

A similar situation has
 emerged in the
heavy-fermion
system
CeCu$_{2}$Si$_{2}$
 (Ref.~\onlinecite{Pang}).
A host of properties, including 
the
inelastic neutron-scattering 
spectrum
~\cite{Stockert_Nat_Phys_2011},
have traditionally been interpreted in terms of
 a sign-changing $d$-wave
 pairing state, yet
recent
specific
heat~\cite{Kittaka} and London penetration
depth~\cite{Pang, Yamashita} 
results 
at very low temperatures
pointed toward a 
fully gapped Fermi-surface
(FS).

It is surprising that 
the
SC phases exhibit $s$- and $d$-wave characters simultaneously. 
One possible origin is $s+\text{i}d$ pairing, which breaks the point-group (PG) and time-reversal (TR) symmetries.
While TR symmetry breaking may develop in special instances in the bulk~\cite{Grinenko_Nat_Phys_2020} or
on the surface~\cite{Zaki_arxiv_2019}, FeSCs typically preserve TRS.
Especially for the alkaline Fe-selenides, there is no evidence for  
either TR symmetry-breaking
or two-stage phase
 transitions as the temperature is lowered. Thus, it is important to identify
an alternate candidate pairing.
For the Fe-chalcogenide SC, 
one candidate pairing state was named $s \tau_3$ 
\cite{Nica_Yu}. 
It has the $s-$wave form factor, but $\tau_{3}$, a
Pauli matrix in the $xz,yz$ $3d$-electron orbital basis, 
turns the pairing state
 into $d-$wave-like; indeed, in the band basis,
 the $ s\tau_3$ pairing
has the 
intra- and inter-band  $d+d $
form. That both intra- and inter-band pairings can play a role is to be expected in this type of model~\cite{Goswami}
and other cases~\cite{Wang_PRB_2010,Ramires_2018}. However,
this $d+d$ form
is highly unusual, thereby raising the question of both its naturalness and generality.

With the stage set by the above, the present work makes two advances. First, we demonstrate
 that 
the $d+d$ pairing state belongs to
matrix
singlet pairing order parameters
with non-trivial orbital structure 
that are
natural and likely common-place
 in multi- orbital/band systems.
 As the orbital degrees-of-freedom (DOF) 
transform non-trivially under PG operations, these matrices can be chosen as one of the irreducible representations 
of the same group. 
 
We make 
a 
case 
for the 
matrix
singlet pairing's naturalness
by presenting an in-depth 
analysis
of the $s\tau_3$ pairing state.
Written in the band basis,
the $s\tau_3$ pairing
has the 
intra- and inter-band  $d+d $
form,
but  
it remains
  an irreducible $B_{1g}$ representation of the
(tetragonal $D_{4h}$) PG.
 The unusual $d+d$ pairing state 
  is to be contrasted with
 its more commonly discussed $d+\text{i}d$ counterpart.
Nonetheless, it is well defined. We demonstrate this point by 
showing that the $d+d$ singlet pairing state can be compared and contrasted with
the more familiar $d+\text{i}d$ state in analogy with how,
in the case of  superfluid \enhe,
the well-defined B-phase
is measured against the equally well-known A-phase.
The latter are spin-triplet pairing states that have
an inherent matrix structure -- in spin space -- even for single-band cases.
 
Second, we
illustrate the
matrix singlet pairing's 
 generality 
 by constructing this type of state
in 
other
multiband systems,
for
 the case of heavy fermion superconductor CeCu$_{2}$Si$_{2}$. 
 This is an important undertaking
 , given that CeCu$_{2}$Si$_{2}$
 is the first-ever discovered unconventional SC \cite{Steg_1979},
 and also recognizing that heavy fermion systems represent a prototype
 setting for strong correlations and unconventional superconductivity in general
\cite{Mathur98,Gegenwart.08,Thompson.12}.
Using  DOF
 that incorporate spin-orbit coupling
(SOC), 
we
introduce
an $s \mathit{\Gamma}_3$ state.
This provides
 the theoretical
  basis for the excellent description of the
experimental results in CeCu$_{2}$Si$_{2}$ in terms of the 
$d+d$ pairing 
order parameter
\cite{Pang, Smidman}.

 We will for the most part direct our analysis towards the effect of multiple orbitals/bands on the nature of the pairing states.
 Therefore, the issue of what drives such pairing states in the multi- orbital/band settings will only be briefly considered. 
 Where this is done, our emphasis is on the 
 short-range spin exchange interactions that are themselves induced by the underlying Coulomb (Hubbard and Hund's) interactions.
What we have achieved, from these microscopic calculations, is to demonstrate the 
 relevance of the general considerations given above.
 We expect that our calculation will motivate further microscopic studies that include additional microscopic physics,
 such as orbital fluctuations,
 or are based on other approaches to the electron-correlation effects.

The emphasis of the present work is on singlet pairing states.
Triplet pairing already has a matrix form that transforms nontrivially in spin space, 
even for single-band systems such 
 as \enhe. However, candidate solid state systems for triplet pairing often involve 
 multiple 
 orbitals/bands
  and strong correlations
 \cite{Ran684,Pustogow2019,Ishida1998,Mack_rmp2003,Rice_1995,Kallin_2009}.
Thus, the type of matrix pairing structure in the orbital/band space we consider here 
 may produce triplet
  superconducting states \cite{Ramires_ruthenate2019,Huang2019}
  and the associated excitations that are of 
 potential
 interest to quantum computing.

The remainder of the paper is organized as follows.
We begin subsection
~\ref{Sec:A}
of our results 
by discussing some of the most relevant general properties of non-trivial matrix pairing 
in the context of the Fe-based SCs. We subsequently define the $s \tau_{3}$ pairing, 
and 
 discuss the unusual properties of this state and show how it can be stabilized in an $s$- to $d$-wave degeneracy  regime .
 We also support our discussion with numerical results for 
 the pertinent five-orbital models of
the Fe-based SCs. Furthermore
, we consider
\enst
in the band basis and 
illustrate how it is analogous to \enhe-B. 
In subsection
~\ref{Sec:B},
we contrast the multiband $d+d$ intra- and inter-band pairing state with the single-band
$d + i d$ pairing state, and argue that these two cases are the analogues of \enhe B and A. 
We show how they can be stabilized in a $t-J_{1}-J_{2}$ model.
In subsection
~\ref{Sec:C} 
, we extend the notion of non-trivial orbital structure beyond the Fe-based compounds by discussing 
a candidate analogous to $s\tau_{3}$ for the heavy-fermion SC CeCu$_{2}$Si$_{2}$.
In order to clearly see these results, we also
present
 the irreducible representations of the $D_{4h}$ point group in the context of CeCu$_{2}$Si$_{2}$. 
The Methods section contains
additional accounts of the numerical results which support the stability of \enst pairing for the alkaline Fe-selenides. We also discuss the $t-J$ model and its solutions which illustrate the case of $d+\text{i}d$ pairing. Additional important aspects of matrix pairing are discussed in the Supplementary 
Notes. 
There, for completeness, we outline the role of the matrix-pairing functions in the various phases of superfluid \enhe, where spin provides the analogue of the orbital DOF. We highlight the lessons we believe can be applied to the case of multiorbital pairing in unconventional SCs. In addition, we illustrate how $s$- and $d$-wave states can coexist without breaking either PG or TR-symmetries via a general Landau-Ginzburg analysis. The band-basis representation of \enst pairing 
and an illustration of the effects of damping on Bogoliubov-de Gennes (BdG) quasiparticles are also
presented in the Supplementary Notes.
 
\section*{Results}

\subsection{$d+d$ matrix singlet pairing as an analogue of \enhe -B} \label{Sec:A}

In solid-state systems, electrons inherit the orbital structure of the underlying ions which form the crystalline lattice. The set of local DOF must include the additional orbital structure. In turn, Cooper pairs formed out of the same electrons are naturally characterized by these additional local, orbital DOF. 

Consider the concrete case of an electronic system on a lattice with $D_{4h}$ tetragonal point-group symmetry. 
Further assume that the dominant contribution to the lowest-lying bands is due to $xz$ and $yz$ orbital local DOF. 
For simplicity, we ignore
SOC. 
  In general, the pairing interactions $V(\mathbf{k}, \mathbf{k'})_{\alpha \beta ;\mathit{\Gamma} \delta}$ 
  depend on the momenta as well as the orbital and spin indices of the two electrons. 
  This two-dimensional space turns out to be 
 relevant for
  Fe-based SCs
  ~\cite{Raghu,Si_Abrahams2008,Daghofer2010},
  and we will first define the $s\tau_3$ pairing state in this space.
  The pairing is orbitally selective in that it is intraorbital and its amplitude is orbital dependent.
We will then consider the stability of the  matrix singlet pairing state
in more realistic five-orbital models.
Through the $d+d$ representation in the band basis, we present an intriguing analogy of the singlet pairing state
as an analogue of \enhe -B. 

\emph{Matrix 
pairing in multiorbital Fe-based SCs}:
 A spin-singlet pairing restricted in the orbital space to the $xz, yz$ cubic harmonics will have the general form 

\enni \begin{align}
\hat{\mathit{\mathit{\Delta}}}(\mathbf{k}) = \mathit{\Delta} \hat{g}(\mathbf{k})_{\alpha \beta} i \sigma_{2}.
\label{Eq:Gnrl_prng_1}
\end{align}

\enni The even-parity matrix $\hat{g}(\mathbf{k})$ denotes the components of the pairing in the four-dimensional space spanned by the tensor products of the two orbital DOF. These tensor-product states are analogues to the spin-1/2 product states in triplet \enhe 
(see Supplementary Note 1). 
Likewise, they depend on the relative 
momentum
of the pair. 
Finally, $i\sigma_{2}$ denotes spin-singlet pairing. We do not consider this additional structure 
since it plays no essential role in the subsequent discussion.     

The pairing matrix can be decomposed into components which transform according to one of the five even-parity irreducible representations of the $D_{4h}$ point group. This allows for additional separation of the DOF as 

\enni \begin{align}
\hat{g}^{(i)}(\mathbf{k})_{\alpha \beta} = g^{(i)}(\mathbf{k}) \hat{\tau}^{(i)}_{\alpha \beta}, 
\label{Eq:Gnrl_prng_2}
\end{align}

\enni where $i$ labels one of the five, even-parity $A_{1g}, A_{2g}, B_{1g}, B_{2g}$ and $E_{g}$ irreducible representations of $D_{4h}$. The functions $g^{(i)}(\mathbf{k})$ can likewise be chosen to belong to one of these representations. To illustrate, $s$-wave states such as $s_{x^{2}+y^{2}}(\mathbf{k})$ and $s_{x^{2}y^{2}}(\mathbf{k})$ both belong to $A_{1g}$. Standard $d$-wave states such as $d_{x^{2}-y^{2}}(\mathbf{k})$ and $d_{xy}(\mathbf{k})$ 
are $B_{1g}$ and $B_{2g}$ representations, respectively. The $xz,yz$ orbital doublet transforms as the two-component $E_{g}$ representation. The $\tau^{(i)}_{\alpha \beta}$ 
identity and Pauli 
matrices describe linear combinations of the tensor-product states which transform according to one of four irreducible representations contained in the $E_{g} \times E_{g} = A_{1g} + A_{2g} + B_{1g} + B_{2g} $ decomposition of the tensor-product space of the two $E_{g}$ orbital DOF. By analogy to the total $S=1$ spin states of \enhe, these matrix-elements play the role of effective Clebsch-Gordan coefficients. The $\tau_{0}, \tau_{1}$, and $\tau_{3}$ matrices transform according to $A_{1g}, B_{2g}$, and $B_{1g}$, respectively. 
In this work, we consider parity-even spin-singlet pairings belonging to one-dimensional irreducible representations of $D_{4h}$. 
This naturally excludes pairing states involving $\tau_{2}$ matrices,
which would be parity-odd.

These arguments point to an important aspect. In \enhe, the relative angular momentum and local (spin) DOF transform independently under separate groups. In the present case, 
$g^{(i)}(\mathbf{k})$
and orbital matrix parts
$(\tau^{(i)})$
are necessarily coupled since they both transform under the same PG. In effect, this constitutes an inherent SOC-like locking of the different spatial DOF of the Cooper pair.  

We note that the single-component representation pairings in Eqs.~\ref{Eq:Gnrl_prng_1} and \ref{Eq:Gnrl_prng_2} are unitary such that

\enni \begin{align}
\hat{\mathit{\Delta}}^{\dag}(\mathbf{k}) \hat{\mathit{\Delta}}(\mathbf{k}) = & \mathit{\Delta}^{2} g^{2}(\mathbf{k}) \tau_{0}.
\end{align}

 Of particular relevance to our discussion is the fact that 
pairing with non-trivial matrix structure
 in general allows for several inequivalent representations of the PG. 
 The problem of determining the stability of the different pairings, including those with non-trivial structure, is 
a challenging task,
which is typically treated numerically on a case-by-case basis. 
We illustrate this point further below in this section, within a five-orbital $t$-$J_1$-$J_2$ model.

\emph{$s\tau_{3}$ pairing state:}
Of interest here is the
\enst pairing. In terms of Eqs.~\ref{Eq:Gnrl_prng_1} and~\ref{Eq:Gnrl_prng_2},
 it corresponds to 
\enni \begin{align}
g(\mathbf{k})= & s_{x^{2}y^{2}}(\mathbf{k}) \\
\hat{\tau}_{\alpha \beta}= & \tau_{3, \alpha \beta}.
\end{align} 

It
 transforms as the $B_{1g}$ representation due exclusively to the $\tau_{3}$ matrix. 
 Because of the orbital struture, it
 represents neither simple $s$- nor $d$-wave states.
 However, \enst pairing preserves both PG- and TR-symmetries of the normal state.

To illustrate the properties of
the
 \enst 
pairing, we first consider a simplified two-orbital model~\cite{Raghu} 
and neglect any possible subleading channels. The TB and pairing parts 
of the BdG Hamiltonian
in the orbital basis
read~\cite{Goswami}

\enni \begin{align}
\hat{H}_{\text{BdG}}(\mathbf{k}) = & \hat{H}_{\text{TB}}(\mathbf{k}) + \hat{H}_{\text{Pair}}(\mathbf{k}) \\ 
\hat{H}_{\text{TB}}(\mathbf{k}) = & \left[ \left(\xi_{0}(\mathbf{k}) - \mu\right) \tau_{0} + \xi_{1}(\mathbf{k}) \tau_{1} + \xi_{3}(\mathbf{k}) \tau_{3} \right] \otimes \gamma_{3} \label{Eq:TB}\\
\hat{H}_{\text{Pair}}(\mathbf{k}) = & \mathit{\Delta}(k) s_{x^{2}y^{2}}(\enht) \tau_{3} \otimes \gamma_{1}.
\end{align}

\enni The $\gamma$ Pauli matrices act in Nambu space. To simplify the expressions, we discuss one of the two spin-sectors. With singlet pairing, the Hamiltonian for the other sector can be obtained in straightforward fashion. Note that, from the perspective of point-group symmetry classification, $s \tau_{3}$ transforms in the same $B_{1g}$ representation 
as the diagonal-in-orbital-space $d_{x^2-y^2} \tau_{0}$ pairing, as has been discussed in this type of model~\cite{Goswami}
and related settings~\cite{Zhou_PRB_2008,Nicholson_PRB_2012, Lv_PRB_2013, Graser.2009}. What distinguishes the $s \tau_{3}$ pairing 
is the nontrivial commutation relation between the corresponding pairing and kinetic parts of the Hamiltonian
~\cite{Nica_Yu}.

It is instructive to recognize that
 the $\xi_{1}\tau_{1}$ and $\xi_{3}\tau_{3}$ terms of $\hat{H}_{\text{TB}}$ play a role similar to a Rashba SOC. The bands corresponding to the normal-state dispersion are 

\enni \begin{align}
\epsilon_{\pm}(\mathbf{k}) = \xi_{0}(\mathbf{k}) \pm \sqrt{\xi^{2}_{1}(\mathbf{k}) + \xi^{2}_{3}(\mathbf{k})},
\label{Eq:Bnds}
\end{align}

\enni reflecting the space-group allowed varying orbital-content and 
splitting of the Fermi surfaces (FSs). We refer the reader to subsection~\ref{Sec:Appn_C} of the Methods for detailed expressions of the $\xi$'s . 
The FS corresponding to this effective model has electron pockets centered at the $(\pm \uppi,0)$ and $(0, \pm \uppi)$ 
points of an one-Fe unit cell.

In Ref.~\onlinecite{Nica_Yu}, we showed that the general BdG dispersion
is \emph{always gapped along the FS}.
Nodes away from the FS can appear for larger band splitting
~\cite{Nica_Yu,Chubukov_PRB_2016, Agterberg_PRL_2017},
reflecting the corrections to the gap term due to the non-commuting TB and pairing parts. 
However, in alkaline-Fe selenides, where \enst pairing is competitive, the 
small band splitting 
at the center of the Brillouin zone
precludes the appearance of nodes.
Even in the cases when the nodes were to appear in the BdG spectrum away from the Fermi surface,
it will not affect our conclusion. 
The point is that, in strongly correlated systems, only nodal excitations on the Fermi energy are long lived and, 
thus, sharply defined. For states away from the Fermi energy, any putative nodal excitations will necessarily 
involve a large damping caused by the underlying electron correlations, 
which obviates the distinction between nodal and gapped excitations. We illustrate how this can occur in Supplementary Note 5.

Another
important characteristic of such a gapped $s\tau_{3}$ state
is its sign change
 under a $\uppi/2$ rotation.
 Such a sign-change leads to the formation of an in-gap spin resonance. 
 $s\tau_{3}$ is then a pairing state which reconciles a fully-gapped FS with the presence of a spin-resonance, 
 typically associated with a $d-$wave gapless order parameter. 

Although we focus on a simplified two-orbital model in order to illustrate the salient properties of \enst pairing, 
the latter can 
also 
be
stabilized in
 more general 
five-orbital
models 
of the alkaline Fe-selenide class of SCs. 
The pairing matrix in the $t-J_{1}-J_{2}$ model can be decomposed into all the symmetry-allowed channels. The complex coefficients of these components have both amplitude and phase.
The illustrative results are shown in
Fig.~\ref{Fig_2}(a,b).
The zero-temperature pairing amplitudes of all symmetry-allowed pairing channels have been 
determined 
in a five-orbital $t-J_{1}-J_{2}$ model with nearest and next-nearest exchange couplings.
This model and
its solution method are discussed in 
subsection~\ref{Appn:5_orb_tJ} of the Methods section.

The TB part and the associated FS are chosen to be consistent with LDA studies~\cite{Nica_Yu}.
The dominant pairing amplitudes are intra-orbital.
The pairing state is orbital selective in the sense that the pairing amplitude and its phase
are orbital sensitive.
 We focus on the case where the pairing 
 amplitude is largest for the $xz,yz$ orbitals 
while also allowing inter-orbital pairing. This reflects orbital-selective correlation effects in the normal state.
The $J_{1}/J_{2}$ ratio 
controls the symmetry of the dominant pairing channel with $s_{x^{2}y^{2}}(\mathbf{k}) \tau_{0}$ and $d_{x^{2}-y^{2}}(\mathbf{k})\tau_{0}$ 
states for small and large values of the ratio, respectively. 
The
\enst
pairing is dominant near the  transition separating order-parameters belonging to $A_{1g}$ to $B_{1g}$ representations for a finite range of the control parameter about the point where $J_{1}/J_{2} \approx 1$.
 A  $d_{x^{2}-y^{2}} \tau_{0}$ with trivial orbital structure provides the subleading pairing with comparable amplitude. 
See subsection~\ref{Sec:Appn_C} 
of the Methods
for more details.

It is important to 
put the results of microscopic studies in a more general perspective.
Our calculations indicate that a subleading $d_{x^{2}-y^{2}}\tau_{0}$ pairing of comparable amplitude 
is present in the regime where \enst is dominant. While we have focused on the properties of the dominant \enst 
pairing alone, 
a more realistic picture would involve coexisting \enst and $d_{x^{2}-y^{2}}\tau_{0}$ in the vicinity of $s$- to $d$-wave 
phase transition. 
This superposition of pairing states with different orbital structure which belong to the same $B_{1g}$ irreducible 
representation preserves both PG- and TR-symmetries.
In Supplementary Note 3
, we present a Landau-Ginzburg analysis to show that, generically,
the pairing state involves a linear superposition of these two components and there is only one superconducting 
transition at a single
$T_c$.

\emph{
Intra- and inter-band $d+d$ pairing and its
analogy 
with the
 \enhe 
 B-phase
}
: It is instructive to consider the $s\tau_{3}$ pairing in a band basis: 

\noindent 

\begin{align}
\label{Eq:Pairing_band}
\hat{H}_{\text{Pair}}(\mathbf{k})=
\mathit{\Delta}_3(\mathbf{k})
 \alpha_3 + 
\mathit{\Delta}_1 (\mathbf{k})
 \alpha_1,
\end{align}

\enni where $\alpha_{1,3}$ are Pauli matrices in the two-band space and 
where the form factors 
$\mathit{\Delta}_{3,1}$ 
transform as $d_{x^{2}-y^{2}}$ and $d_{xy}$, respectively. The details of the transformation from orbital to band basis are discussed in Supplementary Note 4. There, we also show that the 
$\alpha_{3,1}$ matrices are equivalent to $A_{1g}$ and $A_{2g}$ representations, respectively, by applying the inverse transformation from band to orbital basis. The same conclusion can be reached by requiring that each of the two terms in Eq.~\ref{Eq:Pairing_band} transforms as $B_{1g}$.  
Because the overall pairing is in the  irreducible $B_{1g}$ channel, it is natural that the 
intraband $\alpha_{3}$ part has the $d_{x^2-y^2}$ form factor.
Likewise,  the interband $\alpha_{1}$ component has the $d_{xy}$ form factor.
Thus, the \enst pairing is equivalent to a $d+d$ pairing.

When the pairing matrix is squared, the intra- and inter-band $d$-waves add in quadrature as $\mathit{\Delta}^{2}_1(\mathbf{k}) 
+ \mathit{\Delta}^{2}_2 (\mathbf{k})$ to produce a gap which does not close along the FSs centered on the BZ center 
corresponding to the two bands, as shown in Fig.~\ref{Fig:Schm}. 
This is due to the anti-commuting nature of the 
two Pauli matrices $\alpha_{3}$ and $\alpha_{1}$ which denote intra- and inter-band pairing, 
respectively. As in the orbital basis, corrections to this gap are present due to the splitting of the FSs.
 As discussed in the previous subsection, in cases
relevant to 
our
 discussion, these additional effects are typically small and consequently do not close the gap;
and generically, the FS is always fully gapped.
 
The band basis 
reveals a pairing structure which is very similar to that in \enhe -B. Referring to  Supplementary Notes 1 and 2
, the matrix order-parameter in that case is typically expressed as $\hat{\mathit{\Delta}}_{^{3}\text{He - B}}(\mathbf{k}) \sim (\mathbf{k} \cdot \mathbf{\upsigma}) i\sigma_{2}$. This amounts to a linear superposition of $p$-wave 
states, $p_x$, $p_y$ and $p_z$,
 together with a matrix structure made possible due to spin-triplet pairing as represented by the $\mathbf{\upsigma}$ Pauli matrices. The anti-commuting nature of these matrices ensures that three $p$-waves add in quadrature to produce a full gap. The situation clearly mirrors the case of \enst in the band basis, where two $d$-wave states likewise produce a finite gap. The \enst pairing thus provides a remarkable example where a phase which is similar to \enhe -B via a structure in the band-basis is stabilized in a solid-state SC model.

Along with this
similarity between 
the \enst pairing state
and 
the B phase of \enhe, it is important to also 
note on
 the ways in which they differ. 
The distinctions are due primarily to the continuous rotation symmetries of \enhe as contrasted with the discrete nature of the PG in the inter- and intra-band $d$-wave case. The latter belong to a single irreducible representation of a PG involving discrete operations. As such, they break no symmetries of the normal state with the trivial exception of a global phase rotation due to pairing. By contrast, \enhe -B breaks the  $SO(3)_{\mathbf{L}} \times SO(3)_{\mathbf{S}}$ symmetry of the normal state down to $SO(3)_{\mathbf{L}+\mathbf{S}}$, via a relative spin-orbit symmetry breaking~\cite{Leggett, Vollhardt}
. Specifically, the invariance of the normal state under \emph{continuous and independent} rotations of angular-momentum and spin , respectively, is broken down to simultaneous rotations in both sectors. In spite of this additional symmetry-breaking, we note that the B phase has the largest residual symmetry of all superfluid \enhe phases. In this respect, it still resembles intra- and inter-band $d$-wave pairing which preserves both PG and TR symmetries. 

\subsection{
$d+d$ 
and
 $d+\text{i}d$ pairing: Analogy with \enhe -B {\it vs.} \enhe -A
 \label{Sec:B}}

We have seen that an orbital
basis is 
convenient for classifying the pairing states according to symmetry 
and for solving microscopic models.
 Physically, the equivalent band basis is more natural in 
connecting with experiment.
We have also seen that 
the non-trivial \enst pairing is equivalent to simultaneous intra- and inter-band $d_{x^{2}-y^{2}}$ and $d_{xy}$ pairings (Eq.~\ref{Eq:Pairing_band}). These add in quadrature to produce a full gap 
on the FS 
on either of the two bands and their sign-changing factors allow for the formation of in-gap spin resonances. 
For simplicity, we consider only unitary pairings.
The intra- and inter-band terms are consequently associated with $\alpha_{3}$ and $\alpha_{1}$ Pauli matrices, respectively. Importantly, we assume that a $d+d$ pairing does not break either PG or TR symmetries of the Hamiltonian. This amounts to associating both $d$-wave components with a single irreducible representation in an orbital basis.

We have shown that the $d+d$ pairing is a well-defined pairing state, through an analogy with the B phase of \enhe. To further 
elucidate the naturalness of this unusual pairing state, we compare and contrast it with the more familiar $d+\text{i}d$ pairing. We show that 
$d+d$ {\it vs.} $d+\text{i}d$ pairing is analogous to the B- {\it vs.} A-phases of \enhe.

\emph{
$d+d$ in a multiorbital model {\it vs.} $d+\text{i}d$ in a single-orbital model
}
: An \emph{intra-band} $d+\text{i}d$ pairing, where the two components are $d_{x^{2}-y^{2}}$- and $d_{xy}$-waves, respectively, is a natural competitor to the intra- and inter-band $d+d$. 
Here, we show how the intra-band $d+\text{i}d$ can be stabilized in a
one-band
 $t-J_{1}-J_{2}$ model in the vicinity of the $J_{1}\approx J_{2}$ point.

In subsection~\ref{Sec:A}
, we discussed how the \enst orbital non-trivial pairing channel becomes dominant for a finite range of the $J_{1}/J_{2}$ tuning parameter in a realistic five-orbital $t-J_{1}-J_{2}$ model for the alkaline Fe-selenides. The details of the calculations are given in the Methods section. We showed how \enst pairing is equivalent to a $d+d$ intra- and inter-band pairing. To further understand the nature of the \enst -dominated state, we plot the phases of the leading $B_{1g}$ channels relative to \enst as functions of $J_{1}/J_{2}$  
in Fig.~\ref{Fig_2}~(b). The leading $B_{1g}$ channels have relative phases wrt \enst which are closely centered around 0 or $\uppi$ for the entire range of the tuning parameter. In the $[0.8,1]$ interval where \enst is dominant the relative phases are 
either zero or $\pm \uppi$. 
Here the amplitudes of the subleading $d_{x^{2}-y^{2}}\tau_{0}$ and $s_{x^{2}+y^{2}}\tau_{3}$ $B_{1g}$ channels are comparable to that of the leading \enst. Therefore, this regime corresponds to a pairing state with non-trivial orbital structure which  preserves TR and PG symmetries. We note that all $A_{1g}$ and $B_{2g}$ channels are strongly suppressed in the regime where \enst is dominant which we consider here (Fig.~\ref{Fig_5}). 

We next discuss the orbital-trivial $d+\text{i}d$
pairing. For simplicity, we consider a single $d_{xy}$ orbital $t-J_{1}-J_{2}$ model on a square lattice. We choose the tight-binding (TB) parameters and chemical potential to be consistent with a circular hole pocket at the center of the BZ. The details of the model are discussed in subsection~\ref{Sec:Appn_D} of the Methods. The model is solved using a 
self-consistent  
decomposition of the exchange interactions as in the five-orbital cases discussed previously~\cite{Yu, Nica_Yu}. The resulting zero-temperature pairing amplitudes for $J_{2}=1/2$ in units of the bandwidth, and for a finite range of the ratio $J_{1}/J_{2}$ are shown in Fig.~\ref{Fig_3}~(a). For $J_{1}/J_{2} < 0.8$, the only significant pairing occurs in the $d_{xy}, B_{2g}$ channel. For higher values of $J_{1}/J_{2}$, the amplitude of a $d_{x^{2}-y^{2}}, B_{1g}$ pairing becomes finite. These two remain finite up to $J_{2}/J_{1} \approx 2.1$, where the $d_{xy}$ component vanishes continuously. Beyond this point, two additional $s_{x^{2}+y^{2}}$ and $s_{x^{2}y^{2}}$ order-parameters emerge. 
A similar conclusion has been reached in a related model
\cite{Sachdev}.

To illustrate that the two coexisting $d$-wave components are locked into a $d+\text{i}d$ state, we plot their relative phases mod $\uppi$ in Fig.~\ref{Fig_3}~(b). 
The relative phases are obtained from the difference in the phases of each symmetry-allowed channel 
which are determined from the self-consistent solution.
While these relative phases are essentially arbitrary whenever one of the $d$-waves vanishes, a $\uppi/2$ relative phase is clear in the interval $J_{1}/J_{2} \in [0.8, 2.1]$ where both coexist. Although these results were obtained for a single-orbital model, they do demonstrate how a $d+\text{i}d$ pairing with trivial orbital structure can become stable in similar two-orbital models.  

\emph{
$d+d$ pairing {\it vs.} $d+\text{i}d$: Analogy with
B- {\it vs.} A-phases 
of  \enhe superfluid
}
: In subsection~\ref{Sec:A}
, we 
showed that the $d+d$ pairing is closely analogous to the B phase of \enhe ,
 where
the pairing is a superposition of $p_{x,y,z}$-waves corresponding to equal- and opposite-spin pairing as illustrated by Eqs. 2 and 3 of Supplementary Notes 1 and 2. Just like the B phase, the $d+d$ pairing is an irreducible representation, here of the PG, and preserves the TR symmetry of the normal state by construction. 

By contrast, the 
intra-band
$(d+\text{i}d)\alpha_{0}$ pairing, where 
$\alpha_{0}$ is the identity matrix in the band basis, is 
a linear superposition of two irreducible representations. 

In general, the $\alpha_{0}$ matrix in band space would correspond to an identity $\tau_{0}$ matrix in orbital space. The $d+\text{i}d$ pairing spontaneously breaks both PG and TR symmetries. Therefore, it is a natural analogue of \enhe -A,
which is typically described in terms of equal-spin $p_{x} + ip_{y}$ pairing. This phase spontaneously breaks both rotational 
and TR-symmetries of the normal state as illustrated by Eq.~5  of  Supplementary Note 2. The band and spin matrices in the \enhe -A and $d+\text{i}d$ cases are analogous.

In subsection~\ref{Sec:A}
, we discussed how $d+d$ differs from \enhe-B due mainly to the discrete versus continuous symmetries of the two, respectively.  
This kind of difference also exists
between $d+\text{i}d$ and \enhe -A. The latter spontaneously breaks both angular momentum and spin \emph{continuous} rotational symmetries down to a $U_{L_{z} -\phi} \times U_{S_{z}}$ group of independent rotations in each sector about preferred axes (Supplementary Note 2
).
When dipole-dipole interactions are negligible, the directions of either axes are arbitrarily chosen. By contrast, $d+\text{i}d$ involves a superposition of pairings belonging to two irreducible representations of a discrete PG corresponding to fixed symmetry axes. 
Additionally, in 
\enhe-A, the two components, $p_x$ and $p_y$ are exactly degenerate, and there is only a single transition
temperature $T_c$. By contrast, in $d+\text{i}d$, the two d-components are generically non-degenerate, and two stages of phase
transitions are to be expected 
when the temperature is varied.

In spite of clear differences, the formal similarities between \enhe -B and $d+d$ and likewise between \enhe -A and $d+\text{i}d$,
which are due to the presence of non-trivial matrix structure,
are intriguing.  In this sense, \enhe provides both a well-established 
parallel
and a prototype for the emergence and description of the effects of non-trivial matrix structure in unconventional
singlet
 SCs.
 
Given the venerable status of superfluid \enhe -B, we believe that revealing the above connections 
elevates the status of the $d+d$ spin-singlet pairing state.
In turn, this connection
  motivates the consideration that such $d+d$ spin-singlet pairing  beyond the context
 of Fe-based superconductors. Indeed, this leads us to the second part of 
 our work, which is to propose a microscopic pairing state that is capable of understanding
  the heavy fermion superconductor CeCu$_2$Si$_2$. 

\subsection{
Matrix singlet pairing
with spin-orbit coupling:
CeCu$_{2}$Si$_{2}$
\label{Sec:C}
}

Another class of multiband superconductors arises in heavy fermion systems,
in which quasi-localized $f$ electrons hybridize with dispersive $spd$- conduction ($c$) electrons.
These include CeCu$_{2}$Si$_{2}$,
which is the first-ever discovered unconventional SC
~\cite{Steg_1979} and one of the best-studied heavy-fermion SCs. 
For most of 
its
history, this compound was believed to have a conventional $d$-wave order parameter. 
Such 
a
conclusion has been supported by inelastic neutron scattering experiments which revealed 
a spin-resonance peak in the SC state~\cite{Stockert_Nat_Phys_2011} together with angle-resolved 
resistivity measurements of
the upper critical field $H_{\text{c}2}$ ~\cite{Vieyra},
among others.
Remarkably,  recent measurements of the specific heat~\cite{Kittaka} and London penetration depth~\cite{Pang, Yamashita} 
down to lower temperatures
indicated a fully-gapped SC state. The apparent contradiction between these different experimental probes 
is  reminiscent of the situation in the  Fe-chalcogenide SCs. In those cases, we 
argued that the fully-gapped 
but sign-changing $s\tau_{3}$
provide
 a natural resolution. An analogous proposal for CeCu$_{2}$Si$_{2}$ is clearly of great interest. 
 Note
  that a $d+d$ inter- and intra-band pairing directly inspired from the Fe-based cases 
 provides a good fit to the the superfluid-density and specific-heat results in CeCu$_{2}$Si$_{2}$~\cite{Pang, Smidman}.

\emph{
Objective and outline of the section
}
: Here, we construct the analogue of the $s\tau_{3}$ pairing for CeCu$_{2}$Si$_{2}$.
For reasons that will become clear, we
 shall refer to this state as $s\mathit{\Gamma}_{3}$ to indicate the associated non-trivial pairing matrix, as in the case of the Fe-based SCs.

We consider the pairing between the
 composite heavy quasiparticles in terms of simultaneous $f$-$f$, $f$-$c$ and $c$-$c$ pairing in the original 
 electron basis prior to hybridization.
 Of the three, $f$-$f$ pairing is expected to be the strongest, reflecting the more localized nature of the heavy bands. 
 The albeit weaker $f$-$c$ and $c$-$c$ pairing terms will be important, especially when they 
 are involved in creating a pairing component that opens a gap.
 
 In contrast to the case of the Fe-based SC, an important ingredient for constructing pairing states in a heavy fermion metal
such as CeCu$_{2}$Si$_{2}$ is that SOC plays a $0$th-order role. The local orbital and spin DOFs transform simultaneously under PG operations. This imposes additional constraints on any matrix associated with the local DOF. Due to the large
 SOC, the local $f$-electron manifold splits as a consequence of the crystal field. 
 The resulting multiplets, which are labeled according to the irreducible representations of $D_{4h}$, play a role analogous to that of the $d_{xz/yz}$ orbitals in the Fe-based SC case.

A number of experiments~\cite{Goremychkin, Rueff} as well as LDA+DMFT studies~\cite{Pourovskii} 
have indicated that one of the $\mathit{\Gamma}_{7}$ doublets of the crystal-field split $^{2}F_{5/2}$ local electron
 is the dominant contribution to the heavy FS sheets. 
 The lowest-lying excited states of the $f$ electron are composed of a $\mathit{\Gamma}_{6}$ and another $\mathit{\Gamma}_{7}$ doublet. 
Our analysis will also involve $\mathit{\Gamma}_{6}$ Wannier orbitals of the conduction electron states near the Fermi energy, and these Wannier orbitals will hybridize with the excited $\mathit{\Gamma}_{6}$ $f$-level and thereby makes it a small but nonzero component in the ground-state manifold.

In this subsection,
we will 
use these DOFs to advance a matrix pairing state, $s\mathit{\Gamma}_{3}$, which transforms in $B_{1g}$ under $D_{4h}$.
To clarify the involved DOFs, 
 we also discuss the character table of the $D_{4h}$ point group and its irreducible representations and construct the conduction-electron $\mathit{\Gamma}_{6}$ Wannier orbitals from the Cu $3d$ orbitals.

\emph{
Spin-orbital coupled local states
}
: Our aim is to propose a minimal symmetry-allowed candidate for CeCu$_{2}$Si$_{2}$ which has properties 
similar to that of $s\tau_{3}$ in the Fe-based cases. By construction, such a state must belong to one of the 
single-component double-valued irreducible representations of $D_{4h}$, as required by strong SOC. To illustrate, the even-parity double-valued irreducible representation $\mathit{\Gamma}^{+}_{1}$ 
is the analogue of $A_{1g}$, while $\mathit{\Gamma}^{+}_{3}$ is the analogue of $B_{1g}$. The latter is our prime candidate.

Either $f$ or $c$ 
electrons originate from an odd-spin state and therefore transform as either $\mathit{\Gamma}_{6}$ or $\mathit{\Gamma}_{7}$ representations of the PG.
A minimal structure for combined local orbital-spin DOF is determined by a $2\times 2$ matrix $\mathit{\Sigma}$. 
This matrix must belong to a non-trivial irreducible representation of the 
PG; {\it e.g.},
it must change sign under a $C_{4z}$ rotation. 
To ensure that the rotation properties are determined exclusively by the local orbital-spin DOF, 
the pairing must be a product between the non-trivial orbital-spin matrix and a 
form-factor
 belonging to the identity representation. 
In addition to the matrix structure of the local DOF, the pairing matrix must also incorporate $c,f$ indices. 

Thus, a
 minimal order-parameter is
 a $4 \times 4$ matrix.  We consider singlet, parity-even pairing exclusively. 
 Hence, candidate pairing matrices must 
 be 
 \emph{odd under exchange and TR-invariant}. 
 For simplicity, we restrict our discussion to pairings which are even 
 under $f$-$c$ exchange. This necessarily implies that $\mathit{\Sigma}$ is \emph{anti-symmetric}. 
 Furthermore, as the $\mathit{\Sigma}$ matrix can transform under inversion, 
 we only consider pairings between electrons belonging to irreducible representations of identical parity. 
 Following the notation used previously, possible candidates are chosen to be of the form

\enni \begin{align}
\hat{\mathit{\Delta}}(\mathbf{k}) = \mathit{\Delta} g(\mathbf{k}) \hat{\mathit{\Sigma}} \otimes \hat{\mathit{\Xi}}.
\label{Eq:Gnrl_prng_hf}
\end{align} 

\enni The components of the local-DOF multiplet are determined by the $2 \times 2$ $\mathit{\Sigma}$ matrix 
while the $f,c$ nature of the paired electrons is given by the $2 \times 2$ $\mathit{\Xi}$ matrix. As in the more familiar case of full spin rotational symmetry, 
the matrix elements of the $\mathit{\Sigma}$ matrices are effective Clebsch-Gordan coefficients adapted to the cases 
of discrete PG symmetry~\cite{Koster}. 
 
\emph{
Conventional $B_{1g}$ pairing
}
:We first consider candidates on the $\mathit{\Gamma}^{-}_{7}$ ground-state doublet. The superscript denotes odd parity under inversion. Although these naturally correspond to $f$-$f$ pairing involving the $\mathit{\Gamma}_{7}$ ground-state local multiplet,
 they also cover possible 
 $f$-$c$
  pairings with $c$ electrons which belong to the same representation. In the latter case, the $c$ electrons would correspond to $p$-type linear-superposition of Wannier orbitals. The tensor product of two such doublets decomposes into the irreducible representations of $D_{4h}$ as~\cite{Koster}

\enni \begin{align}
\mathit{\Gamma}^{-}_{7} \times \mathit{\Gamma}^{-}_{7} = \mathit{\Gamma}^{+}_{1} + \mathit{\Gamma}^{+}_{2} + \mathit{\Gamma}^{+}_{5}.
\end{align}

\enni Here, $\mathit{\Gamma}^{+}_{1,2}$ are one-dimensional representations which are analogous to the $A_{1g}$ and $B_{2g}$ in the absence of SOC~\cite{Koster}. The two-dimensional $\mathit{\Gamma}^{+}_{5}$ is analogous to the $xz,yz (E_{g})$ doublet. Following Ref.~\onlinecite{Koster}, the matrices corresponding to each of the three $\mathit{\Gamma}^{+}_{1,2,5}$ representations are:

\enni \begin{align}
\Sigma_{\mathit{\Gamma}_{1}} = & \frac{\text{i}}{\sqrt{2}} \sigma_{2} \\
\Sigma_{\mathit{\Gamma}_{2}} = & -\frac{\text{i}}{\sqrt{2}} \sigma_{1} \\
\Sigma^{(5)}_{\mathit{\Gamma}_{5}, x} = & \frac{\text{i}}{\sqrt{2}} \sigma_{3} \\
\Sigma^{(5)}_{\mathit{\Gamma}_{5},y} = & \frac{1}{\sqrt{2}} \sigma_{0}.
\end{align}

\enni The $\sigma$'s are standard Pauli matrices. Recall that for $f$-$f$ or symmetric $f$-$c$ pairings, 
we require  $\mathit{\Sigma}$ to be anti-symmetric. The only choice is $\mathit{\Sigma}_{\mathit{\Gamma}_{1}}$ which transforms as the trivial representation. 
This matrix is the analogue of simple singlet-pairing in the standard BCS case and is invariant under all PG operations. 
It is clear that $f$-$f$ or symmetric $f$-$c$ singlet pairing on the $\mathit{\Gamma}^{-}_{7}$ manifold 
does not support any non-trivial structure in the local DOF. This contrasts with the Fe-based case, 
where the absence of SOC allowed for all $\tau$ matrices in the $xz,yz$ manifold.

We can construct a standard $d$-wave pairing belonging to the $\mathit{\Gamma}_{3}$ representation which is analogous to a $B_{1g}$ representation without SOC. We do so by choosing $g(\mathbf{k})=d_{x^{2}-y^{2}}(\mathbf{k})$ and $\hat{\mathit{\Sigma}}=\mathit{\Sigma}_{\mathit{\Gamma}_{1}}$ 
in Eq.~\ref{Eq:Gnrl_prng_hf}. Likewise, $\hat{\Xi}$ can be chosen to be proportional to either $\mathit{\Xi}_{1}$ or $(1/2)(\mathit{\Xi}_{0}-\mathit{\Xi}_{3})$, 
where $\mathit{\Xi}_{0}$ and $\mathit{\Xi}_{1,3}$ denote identity and Pauli matrices, respectively. The two cases correspond to $f$-$c$ and $f$-$f$ pairing, respectively.

\emph{
Matrix $B_{1g}$ pairing
}
: We next consider pairing between electrons belonging to distinct
 $\mathit{\Gamma}^{-}_{7}$ and $\mathit{\Gamma}^{-}_{6}$ manifolds. 
This can correspond to $f$-$f$ pairing between electrons belonging to the $\mathit{\Gamma}^{-}_{7}$ ground-state 
and $f$ electrons belonging to the excited $\mathit{\Gamma}^{-}_{6}$ manifolds, respectively. 
Alternately, it can 
denote
$f$-$c$
 pairing between $\mathit{\Gamma}^{-}_{7}$ $f$-electrons and $\mathit{\Gamma}^{-}_{6}$ conduction $c$
 electrons. Further below, we illustrate how intra-unit cell linear combinations of Cu $3d$ states in the presence of SOC can form bases for $\mathit{\Gamma}^{-}_{6}$ conduction electrons. The product states decompose as~\cite{Koster}

\enni \begin{align}
\mathit{\Gamma}^{-}_{6} \times \mathit{\Gamma}^{-}_{7} = & \mathit{\Gamma}^{+}_{3} + \mathit{\Gamma}^{+}_{4} + \mathit{\Gamma}^{+}_{5}.
\end{align}

\enni The corresponding matrices are~\cite{Koster}

\enni \begin{align}
\mathit{\Sigma}_{\mathit{\Gamma}_{3}} = & \frac{\text{i}}{\sqrt{2}} \chi_{2} \\
\mathit{\Sigma}_{\mathit{\Gamma}_{2}} = & -\frac{\text{i}}{\sqrt{2}} \chi_{1} \\
\mathit{\Sigma}_{\mathit{\Gamma}_{5}, x} = & \frac{1}{\sqrt{2}} \chi_{0} \\
\mathit{\Sigma}_{\mathit{\Gamma}_{5}, y} = & \frac{\text{i}}{\sqrt{2}} \chi_{3}.
\end{align}

\enni The $\chi$'s are Pauli matrices. Note however that they represent different DOF and thus transform differently under the PG. Therefore, one should not confuse the meaning of the $\chi$ Pauli matrices defined in this case with those of the $\mathit{\Gamma}_{7}-\mathit{\Gamma}_{7}$ case discussed previously. The only anti-symmetric matrix is $\Sigma_{\mathit{\Gamma}_{3}}$. It transforms as the $\mathit{\Gamma}^{+}_{3}$ irrep of $D_{4h}$ and is an analogue 
of the $\tau_{3}$ matrix in the Fe-based cases. Moreover, this matrix is invariant under TR. We conclude that a counterpart of the $s\tau_{3}$ pairing for CeCu$_{2}$Si$_{2}$ is 
an \emph{s-wave pairing belonging to the sign-changing $\mathit{\Gamma}_3$ representation, or $s\mathit{\Gamma}_{3}$  pairing}:

\enni \begin{align}
\hat{\mathit{\Delta}}(\mathbf{k}) = \mathit{\Delta} s(\mathbf{k}) i \chi_{2} \otimes \hat{\mathit{\Xi}} ,
\end{align} 

\enni where
$s(\mathbf{k})$ corresponds to a 
 $s$-wave 
form factor
which transforms according to the $\mathit{\Gamma}^{+}_{1}$ trivial irrep. $s\mathit{\Gamma}_{3}$ pairing, which involves electrons belonging to different irreducible representations due to the $\mathit{\Gamma}_{3}$ matrix, is necessarily non-local, and thus vanishes when $\mathbf{r}_{\text{Relative}}=0$, 
 where $\mathbf{r}_{\text{Relative}}$ is the distance between two paired electrons. Therefore, we do not restrict the $s$-wave form factor to be of sign-changing form. 
 The form of the $\mathit{\Xi}$ matrix differs depending on either 
  $f$-$f$ 
  or
  $f$-$c$
  pairings. In the 
  $f$-$c$ case it can be chosen to be proportional to a $\mathit{\Xi}_{1}$ Pauli matrix. In the $f-f$ case, 
  it can be made proportional to a $\mathit{\Xi}_{0} - \mathit{\Xi}_{3}$ matrix. 
  In either case, the gap is determined
   by the amplitude and 
   form factor
   only similarly to what happens for \enst.
   In a 
   multiband model of CeCu$_{2}$Si$_{2}$~\cite{Pang}, this pairing produces a full gap. 

On general grounds, the non-trivial $\mathit{\Gamma}_{7}$-$\mathit{\Gamma}_{6}$ pairing in either $f$-$f$ or 
$f$-$c$ cases is likely weaker than the $\mathit{\Gamma}_{7}$-$\mathit{\Gamma}_{7}$ $f$-$f$ pairing .
However, there are cases where such $\mathit{\Gamma}_{7}$-$\mathit{\Gamma}_{6}$ contributions can be important. 
Consider a dominant $\mathit{\Gamma}_{7}$-$\mathit{\Gamma}_{7}$ $f$-$f$ pairing corresponding to a $d$-wave state with nodes 
along the FS. An admixture of non-trivial pairing either from $f$-$c$ or from $f$-$f$ involving the excited local manifold 
can open a gap. 
While we can also consider non-trivial pairing terms in the $c$-$c$ sector, these are expected to be weaker than their $f$-$f$ 
and $f$-$c$ counterparts. Likewise, other candidates with non-trivial orbital-spin structure can be obtained
 if we relax some of our assumptions such as the symmetry of the $f$-$c$ terms under exchange. 
 We reserve a detailed analysis of these cases for future work. 

Our candidate $\mathit{\Gamma}_{7}$-$\mathit{\Gamma}_{6}$, $s\mathit{\Gamma}_{3}$
 pairing 
represents an
 $s\tau_{3}$ analogue for CeCu$_{2}$Si$_{2}$. 
 As in the Fe-based cases, the structure of the local DOF allows a natural interpolation between simple 
 $s-$ and $d-$wave states. 
 Such a pairing can in principle reconcile the difficulties in interpreting the more recent experimental results. 

We note that the $\mathit{\Gamma}^{-}_{6}$ conduction electrons which enter the matrix $B_{1g}$ pairing likely originate from Cu $3d$ orbitals(see below). Indeed, several experiments~\cite{Spille, Yuan, Smidman} have indicated that the strongest suppression of $T_{c}$ occurs upon substituting Cu by non-magnetic impurities. Our matrix $B_{1g}$ pairing candidate, which involves $\mathit{\Gamma}^{-}_{6}$ conduction electrons from $3d$ Cu states, is naturally consistent with these findings.

 Similar to the Fe-chalcogenide case, for unitary pairing we expect the 
 $s\mathit{\Gamma}_3$ pairing in the 
band basis to contain the intraband $\alpha_3$ and interband $\alpha_1$ components. Each must be in $d$-wave state,
with the form factor of the intraband $\alpha_3$ being $d_{x^2-y^2}$. Thus, the $s\mathit{\Gamma}_3$ pairing realizes a $d+d$ 
multiband pairing.
Importantly, 
the $d+d$ pairing does not break either PG or TR symmetries of the Hamiltonian. 

As discussed in subsection~\ref{Sec:A}
, we expect that the 
$s\mathit{\Gamma}_3$ matrix-pairing
will coexist with 
a conventional $d$-wave pairing 
below $T_{c}$ since they both belong to the same
 $\mathit{\Gamma}_{3}$ irreducible representation of $D_{4h}$. The admixture between these will ensure that the SC state preserves both PG and TR-symmetries but also exhibits a gap which is finite everywhere along the FS.

We stress that our analysis is distinguished from the well-known symmetry-based procedure typically  considered in the context of heavy-fermion SCs~\cite{Sigrist}. The latter do not explicitly treat possible non-trivial matrix structures associated with the local DOF. Instead, the order-parameters are generically classified according to the irreducible representations of the various PGs in the context of a LG analysis. In our case, we
have focused
on the non-trivial role of the local DOF.

\emph{
Irreducible representations of $D_{4h}$
}
:To expound on the local DOFs, we now turn to the character table for the double-valued representations of the tetragonal $D_{4h}$ point group, showing it in Table~\ref{Tbl:Chrc_tbl}. 

We follow the conventions of Ref.~\onlinecite{Koster}. $E$ represents the identity, $C_{n}$ are rotations about $z$ by $2\uppi/n$, while $C^{'}_{2}$ and $C^{''}_{2}$ are $\uppi$ rotations about $x,y$ and $(-x,y), (x,y)$ axes, respectively. $I$  denotes inversion. $S_{4}$ indicates a $C_{4}$ rotation about z followed by a reflection in the $xy$ plane perpendicular to this axis. $\sigma_{h}$, $\sigma_{v}$, and $\sigma_{d}$ are reflections through the $xy$, $xz$ and $yz$, and diagonal-$z$ planes, respectively. 

In the presence of SOC, the double-valued, even-parity irreducible representations $\mathit{\Gamma}^{+}_{1-5}$ correspond to the single-valued $A_{1g}, A_{2g}, B_{1g}, B_{2g}$, and $E_{g}$ representations, respectively in the absence of SOC. Likewise, the odd-parity $\mathit{\Gamma}^{-}_{1-5}$ correspond to the $A_{1u}, A_{2u}, B_{1u}, B_{2u}$, and $E_{u}$ representations. $\mathit{\Gamma}^{+}_{6}$ and $\mathit{\Gamma}^{+}_{7}$ transform as spinors and product of spinors and linear combinations of $\mathit{\Gamma}^{+}_{1-4}$ and similarly for the odd-parity $\mathit{\Gamma}^{-}_{6}$ and $\mathit{\Gamma}^{-}_{7}$.

Typical bases for representations relevant to this work are $\mathit{\Gamma}^{+}_{3}: x^{2}-y^{2}$, $\mathit{\Gamma}^{+}_{4}: xy$, $\mathit{\Gamma}^{+}_{5}: xz, yz$. For odd-parity representations we mention $\mathit{\Gamma}^{-}_{1}: (x^{2}-y^{2})xyz$, $\mathit{\Gamma}^{-}_{3}: xyz$, $\mathit{\Gamma}^{-}_{6}: \mathit{\Gamma}^{+}_{6} \times \mathit{\Gamma}^{-}_{1}$ and $\mathit{\Gamma}^{-}_{7}: \mathit{\Gamma}^{+}_{7} \times \mathit{\Gamma}^{-}_{3}$.  

\emph{
$\mathit{\Gamma}^{-}_{6}$ conduction electrons
}
:
Previously, we discussed matrix $B_{1g}$ pairing between $\mathit{\Gamma}^{-}_{6}$ $c$
-electrons and $\mathit{\Gamma}^{-}_{7}$ f-electrons. In this subsection, we illustrate how linear combinations of intra-unit-cell Cu $3d$ orbitals in the presence of SOC provide $\mathit{\Gamma}^{-}_{6}$ conduction electron states. We consider crystal field-split $d_{x^{2}-y^{2}}$ Cu orbitals for simplicity although similar constructions are possible for other orbitals.    

Consider one of the two Cu planes in the unit cell of CeCu$_{2}$Si$_{2}$~\cite{Pourovskii}. The Cu sites are located halfway along the edges of a plaquette as illustrated in Fig.~\ref{Fig_4}.  

\enni The linear combinations of the four orbitals at the Cu sites

\enni \begin{align}
p_{x} = & d^{(4)}_{x^{2}-y^{2}} - 
d^{(2)}_{x^{2}-y^{2}} 
 \\
p_{y} = & d^{(1)}_{x^{2}-y^{2}} - d^{(3)}_{x^{2}-y^{2}} 
\end{align}

\enni transform as an $(x,y)$ doublet under the $D_{4h}$ point group. Consequently, $(p_{x}, p_{y})$ belong to a two-component $\mathit{\Gamma}^{-}_{5}$ irreducible representation. We further include the local spin-1/2 DOF which belongs to a $\mathit{\Gamma}^{+}_{6}$ irreducible representation~\cite{Koster}. From the direct-product states of $p$-orbital linear combinations and spinor states $\phi_{\pm 1/2}$ we can construct states which belong to $\mathit{\Gamma}^{-}_{6}$ representation~\cite{Koster}:

\enni \begin{align}
\Psi_{\mathit{\Gamma}^{-}_{6}; 1/2} = & \frac{\text{i}}{\sqrt{2}} \left[ p_{x}+\text{i}p_{y} \right] \phi_{-1/2} \\
\Psi_{\mathit{\Gamma}^{-}_{6}; -1/2} = & \frac{\text{i}}{\sqrt{2}} \left[ p_{x}-\text{i}p_{y} \right] \phi_{1/2},
\end{align}

\enni where SOC was taken into account. Similar states can be constructed in the remaining Cu plane. The symmetric linear combination between $\mathit{\Gamma}^{-}_{6}$ states in both Cu planes likewise belongs to $\mathit{\Gamma}^{-}_{6}$ doublet. 

\section*{Discussion}
\label{Sec:Cncl}

Recent experiments in multiband Fe-based and heavy-fermion SCs are inconsistent with either simple $s$- or $d$-wave pictures, 
with no conclusive evidence for
{time reversal symmetry breaking.
We argued for alternatives which can interpolate between the two simple cases without breaking the PG and TR 
symmetries via pairings with non-trivial matrix-structure in the orbital DOF. 
We discussed how 
matrix singlet pairings can
emerge in unconventional SCs.

To support our general arguments, we considered the specific context of the Fe-based SCs. 
We present microscopic results showing that the phase difference of the 
intra-band $d_{x^2-y^2}$ and inter-band $d_{xy}$ pairing components 
to be either $0$ or $\uppi$. 
This  $d+d $ pairing is the band basis equivalent of the
 \enst form in the orbital basis,
 and is an irreducible $B_{1g}$ representation of the (tetragonal $D_{4h}$) PG.
 We demonstrate that this  $d+d$  singlet pairing state is well defined,
 by showing that it can be compared and contrasted with the more familiar $d+\text{i}d$ state
in a way analogous to how
the well-defined B-phase in the case of  superfluid \enhe
is measured against the equally well-known A-phase.
The $d+d$ pairing state allows for the reconcillation between seemingly contradictory experimental 
observations.

Non-trivial orbital structure can be relevant to unconventional SCs beyond the Fe-based family. 
To illustrate this, we constructed a pairing analogous to $s\tau_{3}$ for the heavy-fermion CeCu$_{2}$Si$_{2}$ using general group-theoretical arguments. 
This
$s\mathit{\Gamma}_3$
 pairing state is also expected to have a $d+d$ pairing structure in the band basis.
It provides a natural theoretical basis
 to understand the striking
 low-temperature properties 
recently
measured in the superconducting state of
CeCu$_{2}$Si$_{2}$.

In these $d+d$ pairing states, the anti-commuting nature of the two pairing components 
leads to their contributing to the single-particle excitation spectrum 
through an addition in quadrature,
making it a fully-gapped superconducting state.
The formation of the gap is connected to the energetic stabilization of such a state over a range of microscopic parameters.
These results lead us to suggest 
 that $d-$wave superconductors of strongly correlated multiorbital systems will inherently have a 
fully-gapped Fermi surface,
even though the gap can be very small.

{\em Note added}: 
During the reviewing process of this manuscript, Ref.~\onlinecite{Amorese_arxiv_2020} 
appeared with the results of recent x-ray spectroscopy experiments in CeCu$_{2}$Si$_{2}$
that support the $s\mathit{\Gamma}_{3}$ matrix pairing proposed here for CeCu$_{2}$Si$_{2}$.
The $s\mathit{\Gamma}_{3}$ pairing includes paired $\mathit{\Gamma}_{7}$ $f$-electrons and $\mathit{\Gamma}_{6}$ 
conduction electrons. As the latter must hybridize with the excited $\mathit{\Gamma}_{6}$ $f$-electron states,
a small but nonzero mixture of $\mathit{\Gamma}_{6}$ $f$-electrons is expected in the ground-state
manifold. This mixture was shown for CeCu$_{2}$Si$_{2}$ in Ref.~\onlinecite{Amorese_arxiv_2020}.

\setcounter{subsection}{0} 

\section*{Methods}
\label{Sec:Mthd}}

\subsection{Pairing channels of the five-orbital $t-J_{1}-J_{2}$ model}
\label{Sec:Appn_C}

We present our results for the five-orbital $t-J_{1}-J_{2}$ model of the alkaline Fe-selenides \cite{Nica_Yu}. The leading pairing amplitudes at zero-temperature are shown in Fig.~\ref{Fig_5} as functions of $J^{xz/yz}_{1}/J^{xz/yz}_{2}$. Exchange interactions in the $xz/yz$ sector are identical for the two orbitals. $J^{xz/yz}_{2} = 1/2$ in units of the band-width while the exchange interactions for $d_{xy}$ orbital are 5 times smaller. Interactions in the remaining orbitals are ignored. $J_{1}$ and $J_{2}$ refer to their values for the $xz/yz$ sector. For small and large values of the tuning parameter, $s_{x^{2}y^{2}} \tau_{0}, A_{1g}$ and $d_{x^{2}-y^{2}} \tau_{0}, B_{1g}$ orbital-trivial pairings are dominant. In the interval $0.8 \le J_{1} /J_{2} \le 1$, the $s_{x^{2}y^{2}}\tau_{3}$ pairing with non-trivial orbital structure is dominant with sub-leading $d_{x^{2}-y^{2}} \tau_{0}$ channel. 
The abrupt change around $J_{1}/J_{2} \approx 0.75$ is due to a transition from dominant $A_{1g}$ to $B_{1g}$ channels which become quasi-degenerate in this region. For the FS considered here, large $J_{2}$ favors dominant $s_{x2y2}\tau_{0}$, $A_{1g}$ while large $J_{1}$ favors $d_{x2-2y2}\tau_{0}$, $B_{1g}$ channels, respectively. 

The form factors for the pairing terms have the standard expressions

\enni \begin{align}
s_{x^{2}+y^{2}}(\mathbf{k}) = & 
\frac{1}{2}
\left[
\cos(k_{x}a) + \cos(k_{y}a) 
\right]
\\
s_{x^{2}y^{2}}(\mathbf{k}) = & \cos(k_{x}a) \cos(k_{y}a) \\
d_{x^{2}-y^{2}}(\mathbf{k}) = & 
\frac{1}{2}
\left[ \cos(k_{x}a) - \cos(k_{y}a) \right] \\
d_{xy}(\mathbf{k}) = & \sin(k_{x}a) \sin(k_{y}a).
\end{align}

\subsection{Five-orbital $t-J_{1}-J_{2}$ model and solution method}
\label{Appn:5_orb_tJ}

The pairing instabilities in the different symmetry channels of the alkaline Fe-selenides were obtained via a five-orbital $t-J_{1}-J_{2}$ model:

\begin{widetext}
\begin{align}
\label{Eq:t_J_Hamiltonian}
H=  -  & \sum_{i<j} (t_{ij}^{\alpha\beta}c_{\alpha}^{\dagger}c_{\beta} + \text{h.c.}) + \sum_{i,\alpha} \left(  \epsilon_{i\alpha} - \mu \right) n_{i}+  \sum_{\braket{ij},\alpha, \beta}J_{1}^{\alpha\beta} \left( \mathbf{S}_{i\alpha} \cdot \mathbf{S}_{j\beta} - \frac{1}{4}n_{i\alpha}n_{j\beta} \right) +  \notag \\
& + \sum_{\braket{\braket{ij}},\alpha, \beta}J_{2}^{\alpha\beta} \left( \mathbf{S}_{i\alpha} \cdot \mathbf{S}_{j\beta} - \frac{1}{4}n_{i\alpha}n_{j\beta} \right),
\end{align} 
\end{widetext}

\enni where $i,j$ indices cover all of the sites of a two-dimensional square lattice and $\alpha,\beta \in \{1, \hdots 5\}$ represent the $d_{xz}, d_{yz}, d_{x^{2}-y^{2}}, d_{xy}$, and $d_{z^{2}}$ orbitals, respectively. 
The parameters of the model are specified 
 in Ref.~\onlinecite{Yu}. 
Different
orbitals exhibit varying degrees of correlations, such that the exchange couplings are orbital dependent. 
More specifically, intra-orbital exchange couplings for the $d_{x^{2}-y^{2}}$ and $d_{z^{2}}$ orbitals are set to zero. 
Both NN and next-NN (NNN) exchange couplings are equal in the $d_{xz/yz}$ sector and are larger by a factor of 5
 than the exchange couplings in the $d_{xy}$ sector. Inter-orbital exchanges have a small effect~\cite{Nica_Yu} 
 and are neglected here. 

The interactions are decoupled in the particle-particle channel and the model is solved at $T=0$ within a 
self-consistent
 approach. The double-occupancy constraint is introduced via an effective band renormalization~\cite{Yu, Nica_Yu}. 

We calculate the intra-orbital, NN and NNN pairing bonds,  driven by $J_1$ and $J_2$ exchange couplings respectively, along 
$\mathbf{\hat{x}}$, $\mathbf{\hat{y}}$ and $\mathbf{\hat{x}}+\mathbf{\hat{y}}$ and $ \mathbf{\hat{x}}-\mathbf{\hat{y}}$ respectively as
\enni \begin{align}
\mathit{\Delta}_{\mathbf{e}, \alpha \alpha}
= & \frac{1}{2} \braket{
c^{\dag}_{\mathbf{R}_{i} \alpha \up} c^{\dag}_{\mathbf{R}_{i} + \mathbf{e} \alpha \dn} - c^{\dag}_{\mathbf{R}_{i} \alpha \dn} c^{\dag}_{\mathbf{R}_{i} + \mathbf{e} \alpha \up}
}
\end{align}
\enni where $\mathbf{e} \in \{ \mathbf{\hat{x}}, \mathbf{\hat{y}},\mathbf{\hat{x}}+\mathbf{\hat{y}}, \mathbf{\hat{x}}-\mathbf{\hat{y}}\} $, $\mathbf{R}_{i}$ is a site vector, and $\alpha$ is an orbital index. The NN and NNN pairing bonds enter the pairing part of a Nambu Hamiltonian via=
\enni \begin{align}
\mathit{\Delta}_{\mathbf{k}, \alpha \alpha} = & \sum_{\mathbf{e}} J_{\mathbf{e}} \cos \left( \mathbf{k} \cdot \mathbf{e} \right)
\end{align}
The \emph{dimensionless pairing amplitudes} reported in the Results section are obtained by taking appropriate  linear combinations of the NN and NNN pairing bonds. 
As such, the procedure does not bias towards any particular pairing channel. In addition, there are no approximations for the shape of the FS, and the pairing bonds are determined via averages where the momentum summation is over the entire Brillouin zone.
The calculation is initiated with a random set of NN and NNN pairing bonds for all of the orbitals and subsequently allowed to converge. The procedure is repeated until a set of 300 converged solutions are obtained. From this set of converged solutions, we select the one which corresponds to the absolute minimum in the associated free-energy. This solution, again obtained without any superfluous conditions, corresponds to the physical solution reported in the manuscript.   

\subsection{Single-orbital $t-J_{1}-J_{2}$ model}

\label{Sec:Appn_D}

The Hamiltonian of the single $d_{xy}$ orbital $t-J_{1}-J_{2}$ model defined on a 2D square lattice is 

\enni \begin{align}
H =  H_{\text{TB}} + J_{1} \sum_{\braket{ij}} \mathbf{S}_{i} \cdot \mathbf{S}_{j} + J_{2} \sum_{\braket{\braket{ij}}} \mathbf{S}_{i} \cdot \mathbf{S}_{j},
\end{align}

\enni where $i,j$ label the lattice sites. The spin-density operators are defined as $\mathbf{S}_{i} = (1/2) \sum_{ab} c^{\dag}_{a}(\mathbf{R}_{i}) \mathbf{\upsigma}_{ab} c_{b}(\mathbf{R}_{j})$, where $a,b$ are spin indices. 

The TB part is determined by 

\enni \begin{align}
H_{TB} = & -t_{1} \sum_{\braket{ij}} \sum_{a}  c^{\dag}_{a}(\mathbf{R}_{i}) c_{a}(\mathbf{R}_{j}) \notag \\
- & t_{3} \sum_{\braket{\braket{ij}}} \sum_{a}  c^{\dag}_{a}(\mathbf{R}_{i}) c_{a}({\mathbf{R}_{j}}) - \mu \sum_{i} \sum_{a}  c^{\dag}_{a}(\mathbf{R}_{i}) c_{a}(\mathbf{R}_{i}).
\end{align}

\enni The band is determined by 

\enni \begin{align}
\epsilon(\mathbf{k}) = & -2t_{1} \left[ \cos(k_{x}a) + \cos(k_{y}a) \right] \notag \\ 
- & 4t_{3} \cos(k_{x}a)\cos(k_{y}a) - \mu,
\label{Eq:Sngl_orb_TB}
\end{align}

\enni where $a$ is the NN distance. 

The TB coefficients are chosen as $t_{1} = 2t_{3} = -0.5$. The resulting band is shown in Fig.~\ref{Fig_6}. Near half-filling we take 
$\mu \approx -0.3$
to obtain the FS shown in Fig.~\ref{Fig_7}.

We implicitly take into account the  renormalization of the bandwidth near half-filling by considering a large, fixed effective $J_{2}=-2t_{1}=1$ while $J_{1}$ is allowed to vary.
We decouple the exchange interactions in the pairing channels. The model is solved using the methods of Refs.~\onlinecite{Yu, Nica_Yu} near half-filling. 

\section*{Data Availability}

The data that support the findings of this study are available from the corresponding
author upon reasonable request.

\section*{Code Availability}

The codes that support the findings of this study are available from the corresponding
author upon reasonable request.

\section*{Acknowledgements}

We thank 
Pengcheng Dai, J. C. S\'eamus Davis,
 Onur Erten, Haoyu Hu, Andrea Severing, Michael Smidman, 
Frank Steglich, L. Hao Tjeng, Roxanne Tutchton,
Ming Yi,
Rong Yu, Huiqiu Yuan, Jian-Xin Zhu 
and Gertrud Zwicknagl
for useful discussions. 
This work has been supported by ASU startup grant (E.M.N.) 
and 
by the DOE BES Award No. DE-SC0018197 and the Robert A. Welch Foundation Grant No. C-1411 (Q.S.).
We 
acknowledge the hospitality of 
 Center for Nonlinear Studies at Los Alamos National Laboratory, 
 where part of this work was initiated.
Q.S. acknowledges  the support by a Ulam Scholarship
of the Center for Nonlinear Studies 
and
the hospitality of the Aspen Center for Physics (NSF grant No. PHY-1607611).

\section*{Author Contributions}

Both authors contributed equally in the design of this study, in the acquisition and interpretation of the supporting data, and in the drafting of the text. 

\section*{Competing Interests}

The Authors declare no Competing Financial or Non-Financial Interests.


\pagebreak

\section*{Figure Legends}

\setcounter{figure}{0} 

\begin{figure}[h!]
\caption{
\textbf{
Zero-temperature results for a five-orbital $t-J_{1}-J_{2}$ model}
. (a) 
Dimensionless pairing amplitudes of the leading $B_{1g}$ channels as compared
 to that of the $B_{1g}$ \enst channel as functions of $J_{1}/J_{2}$ for a five-orbital $t-J_{1}-J_{2}$ model of the alkaline Fe-selenides. The numerically-determined pairing amplitudes are the weights of each of the PG symmetry-allowed channels.
 $\bm{1}_{xy}$ denotes the trivial $1 \times 1$ matrix in the $d_{xy}$ orbital sector. See subsections A and B of the Methods for the details of the calculation. The \enst pairing with non-trivial orbital strcuture is dominant in the $0.8 \le J_{1}/J_{2} \le 1.0$ $B_{1g}$ window. (b) Phases of the leading $B_{1g}$ channels relative to the \enst channel
 as functions of the tuning parameter. These are obtained from the difference in the  phases of each symmetry-allowed channel which are determined from the self-consistent solution.
 In the $[0.8,1]$ interval where \enst is dominant, these relative phases are 
 either zero or $\pm \uppi$. 
 Here, the amplitudes of the coexisting $B_{1g}$ channels are comparable to that of \enst. This illustrates that the \enst pairing which is equivalent to $d+d$, effectively preserves TR and PG symmetries. 
 }
\end{figure}

\begin{figure}[h!]
\centering
 \caption{
\textbf{
The gapping of an illustrative FS by $d+d$ pairing
}. The blue, dashed line indicates the nodes of the intra-band component $\mathit{\Delta}_{3}$ (Eq.~\ref{Eq:Pairing_band}) , which transforms as a $B_{1g}$ representation of $D_{4h}$. The red, dotted line shows the nodes of the inter-band component $\mathit{\Delta}_{1}$ (also in Eq.\ref{Eq:Pairing_band}), which transforms as a $B_{2g}$ representation. The two components add in quadrature to produce a nonzero gap everywhere on the FS. Note that possible nodes of a common $s$-wave form factor (Eqs.~28-29
 of Supplementary Note 4 for \enst) are not shown here as they are irrelevant to our argument. 
 } 
 \end{figure} 
 
\begin{figure}[h!]
\caption{
\textbf{Zero-temperature leading pairing amplitudes (dimensionless) as functions of 
$J^{xz/yz}_{1}/J^{xz/yz}_{2}$ for a five-orbital $t-J_{1}-J_{2}$ model of the alkaline Fe-selenides.} $J^{xz}_{2}= J^{yz}_{2}=1/2$ in units of the bandwidth. The exchange interactions for the $d_{xy}$ orbital are five times smaller while the exchange couplings for all remaining orbitals are zero. Please see Ref.~\onlinecite{Nica_Yu} for a detailed account of the model and solution. For $J_{1}/J_{2} \le 0.7$ $A_{1g}$ pairing channels are dominant with leading $s_{x^{2}y^{2}} \tau_{0}$ in the $xz/yz$ sector. This pairing has trivial orbital structure. There is a narrow region of coexistence between finite $A_{1g}$ and $B_{1g}$ channels in the $0.7 \le J_{1} / J_{2} \le 0.8$. Beyond this range, $B_{1g}$ channels dominate with leading $s_{x^{2}y^{2}} \tau_{3}$ in the $xz/yz$ sector which has non-trivial orbital structure. At even larger values, an orbital-trivial $d_{x^{2}-y^{2}}\tau_{0}$ phase dominates.
}
\end{figure}

\begin{figure}[h!]
\caption{
\textbf{Zero-temperature results for a single-orbital $t-J_{1}-J_{2}$ model close to half-filling.} Please see subsection C of the Methods for details of the model. (a) Dimensionless pairing amplitudes as functions of the ratio $J_{1}/J_{2}$. When the tuning parameter is less than $0.8$, only the $d_{xy}, B_{2g}$ channel has finite amplitude. In the $0.8 \le J_{1}/J_{2} \le 2.1$ interval, $d_{xy}$ coexists with a $d_{x^{2}-y^{2}}, B_{1g}$ channel. For larger values of the tuning parameter, the $d_{xy}$ channel is suppressed and $s_{x^{2}+y^{2}}$ and $s_{x^{2}y^{2}}$ $A_{1g}$ channels emerge.~(b) Relative phase of the two $d$-wave channels modulo $\uppi$ as a function of $J_{1}/J_{2}$. When both $d$-waves have finite amplitudes, a $\uppi/2$ relative phase is clearly visible. When one of the two is suppressed, the relative phase is essentially arbitrary.
}
\end{figure}

\begin{figure}[h!]
\caption{
\textbf{Single Cu plane in the unit cell of CeCu$_{2}$Si$_{2}$~\cite{Pourovskii}.} The four sites labeled $(1)-(4)$ correspond to Cu $d_{x^{2}-y^{2}}$ orbitals in the plane. The dashed-line circles represent the Ce sites projected onto the Cu-plane.  
}
\end{figure}

\begin{figure}[h!]
\caption{\textbf{Dispersion corresponding to Eq.~\ref{Eq:Sngl_orb_TB} 
in units of $2|t_{1}|$
with $t_{1} = 2t_{3} = -0.5$. }
 $\mu \approx -0.3$ ensures the 
 FS shown in Fig.~\ref{Fig_7} near half-filling. 
}
\end{figure}

\begin{figure}[h!]
\caption{\textbf{Hole-like FS for a single-orbital $d_{xy}$ model close to half-filling.}}
\end{figure}
\pagebreak

\section*{Tables}

\setcounter{table}{0}

\begin{table*}[h!]
\caption{Character table for the double-valued representations of the tetragonal $D_{4h}$ point group}
\begin{ruledtabular}
\begin{tabular}{c c c c c c c c c c c c c c c}
$D_{4h}$ & $E$ & $\bar{E}$ & $2C_{4}$ & $2\bar{C}_{4}$ & $C_{2}/\bar{C}_{2}$ & $2C^{'}_{2}/2\bar{C}^{'}_{2}$ & $2C^{''}_{2}/2\bar{C}^{''}_{2}$ & $I$ & $\bar{I}$ & $2S_{4}$ & $2\bar{S}_{4}$ & $\sigma_{h}/\bar{\sigma}_{h}$ & $2\sigma_{v}$ & $2\sigma_{d}/2\bar{\sigma}_{d}$ \\
\hline 
$\mathit{\Gamma}^{+}_{1} $ & 1 & 1 & 1 & 1 & 1 & 1 & 1 & 1 & 1 & 1 & 1 & 1 & 1 & 1 \\
$\mathit{\Gamma}^{+}_{2} $ & 1 & 1 & 1 & 1 & 1 & $-1$ & $-1$ & 1 & 1 & 1 & 1 & 1 & $-1$ & $-1$ \\
$\mathit{\Gamma}^{+}_{3} $ & 1 & 1 & $-1$ & $-1$ & 1 & 1 & $-1$ & 1 & 1 & $-1$ & $-1$ & 1 & 1 & $-1$ \\
$\mathit{\Gamma}^{+}_{4} $ & 1 & 1 & $-1$ & $-1$ & 1 & $-1$ & 1 & 1 & 1 & $-1$ & $-1$ & 1 & $-1$ & 1 \\
$\mathit{\Gamma}^{+}_{5} $ & 2 & 2 & 0 & 0 & $-2$ & 0 & 0 & 2 & 2 & 0 & 0 & $-2$ & 0 & 0 \\
$\mathit{\Gamma}^{+}_{6} $ & 2 & $-2$ & $\sqrt{2}$ & $-\sqrt{2}$ & 0 & 0 & 0 & 2 & $-2$ & $\sqrt{2}$ & $-\sqrt{2}$ & 0 & 0 & 0 \\
$\mathit{\Gamma}^{+}_{7} $ & 2 & $-2$ & $-\sqrt{2}$ & $\sqrt{2}$ & 0 & 0 & 0 & 2 & $-2$ & $-\sqrt{2}$ & $\sqrt{2}$ & 0 & 0 & 0 \\
$\mathit{\Gamma}^{-}_{1} $ & 1 & 1 & 1 & 1 & 1 & 1 & 1 & $-1$ & $-1$ & $-1$ & $-1$ & $-1$ & $-1$ & $-1$ \\
$\mathit{\Gamma}^{-}_{2} $  & 1 & 1 & 1 & 1 & 1 & $-1$ & $-1$ & $-1$ & $-1$ & $-1$ & $-1$ & $-1$ & 1 & 1 \\
$\mathit{\Gamma}^{-}_{3} $ & 1 & 1 & $-1$ & $-1$ & 1 & 1 & $-1$ & $-1$ & $-1$ & 1 & 1 & $-1$ & $-1$ & 1 \\
$\mathit{\Gamma}^{-}_{4} $ & 1 & 1 & $-1$ & $-1$ & 1 & $-1$ & 1 & $-1$ & $-1$ & 1 & 1 & $-1$ & 1 & $-1$ \\
$\mathit{\Gamma}^{-}_{5} $ & 2 & 2 & 0 & 0 & $-2$ & 0 & 0 & $-2$ & $-2$ & 0 & 0 & 2 & 0 & 0 \\
$\mathit{\Gamma}^{-}_{6} $ & 2 & $-2$ & $\sqrt{2}$ & $-\sqrt{2}$ & 0 & 0 & 0 & $-2$ & 2 & $-\sqrt{2}$ & $\sqrt{2}$ & 0 & 0 & 0 \\
$\mathit{\Gamma}^{-}_{7} $ & 2 & $-2$ & $-\sqrt{2}$ & $\sqrt{2}$ & 0 & 0 & 0 & $-2$ & 2 & $\sqrt{2}$ & $-\sqrt{2}$ & 0 & 0 & 0
\end{tabular}
\end{ruledtabular}
\label{Tbl:Chrc_tbl}
\end{table*}

\pagebreak
\newpage

\section*{Figures}

\setcounter{figure}{0} 

\begin{figure}[h!]
\includegraphics[width=1.0\columnwidth]
{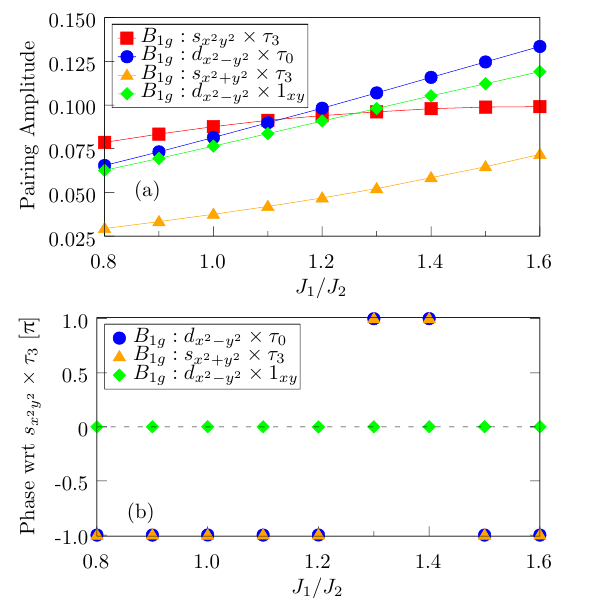}
\caption{}
\label{Fig_2}
\end{figure}
\enni
\begin{figure}[h!]
\centering
\includegraphics[width=1.0\columnwidth]
{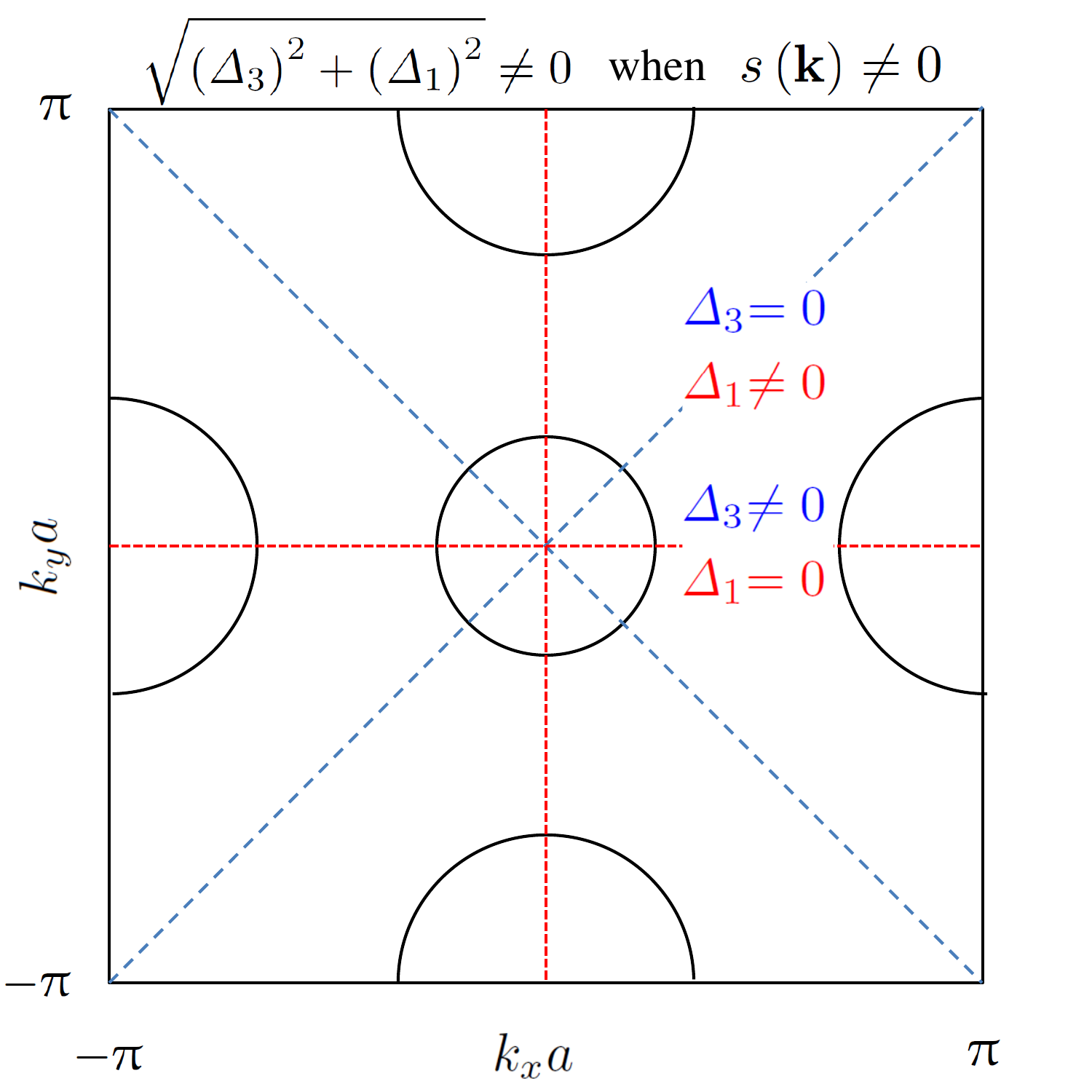}
 \caption{} 
 \label{Fig:Schm}
\end{figure} 
\begin{figure}[h!]
\centering
\includegraphics[width=1.0\columnwidth]
{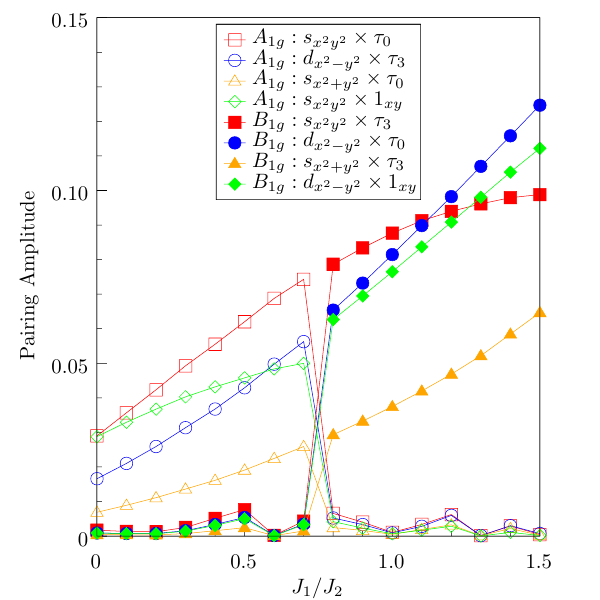}
\caption{}
\label{Fig_5}
\end{figure}
\begin{figure}[h!]
\includegraphics[width=1.0\columnwidth]
{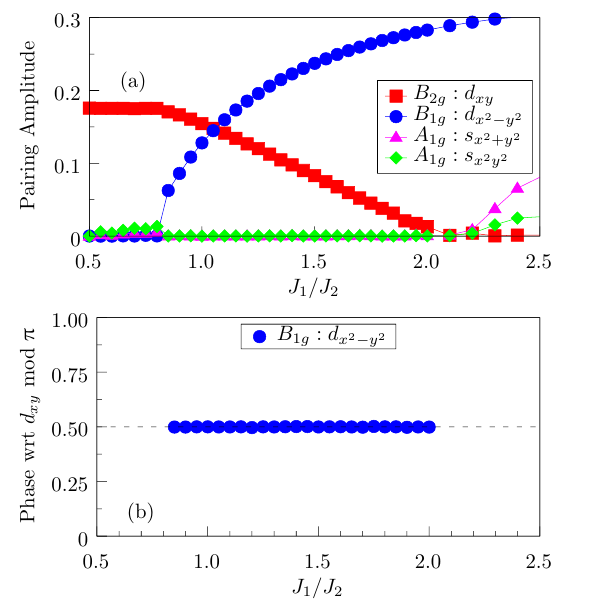}
\caption{}
\label{Fig_3}
\end{figure}
\begin{figure}[h!]
\includegraphics[width=1.0\columnwidth]
{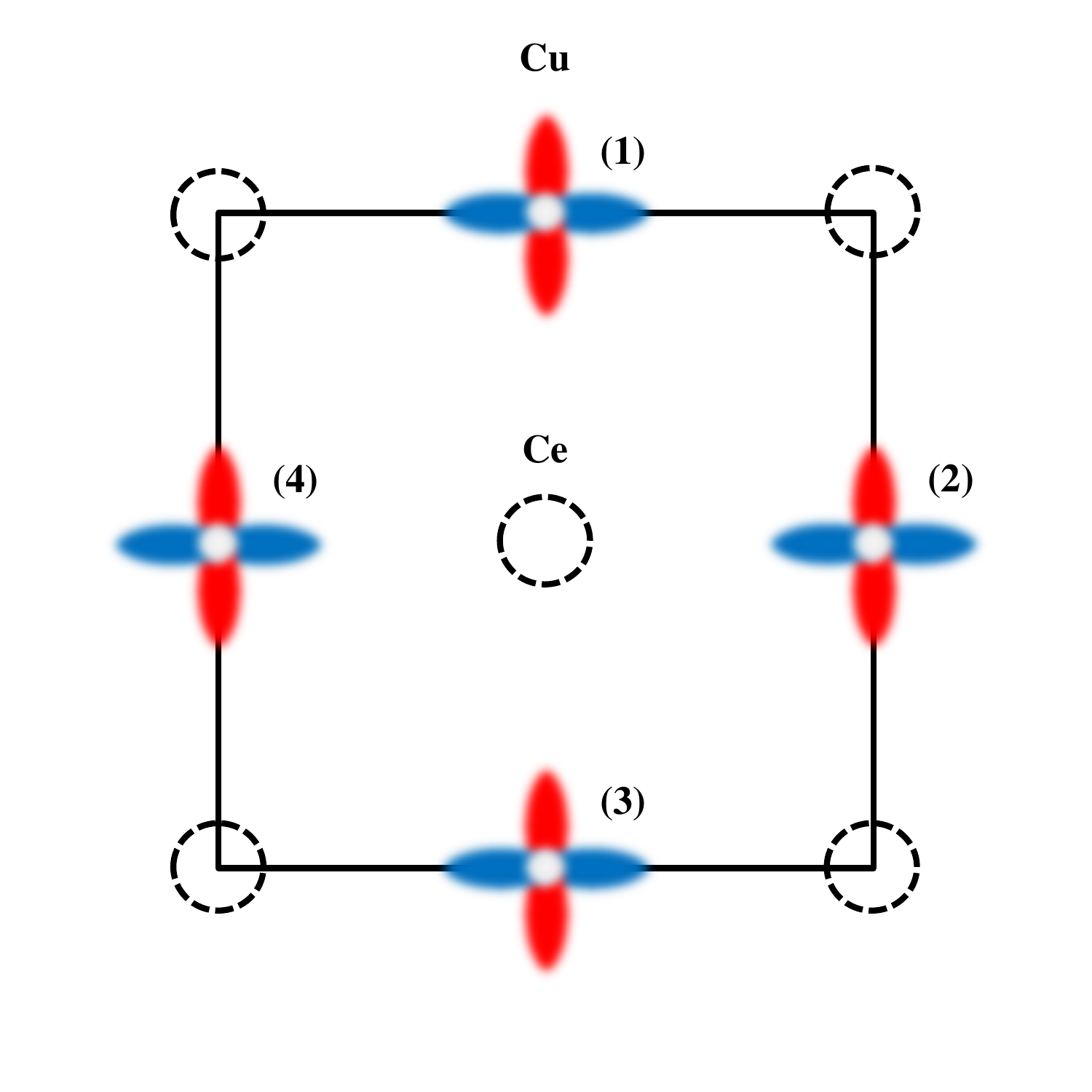}
\caption{}
\label{Fig_4}
\end{figure}
\enni \begin{figure}[h!]
\centering
\includegraphics[width=1.0\columnwidth]{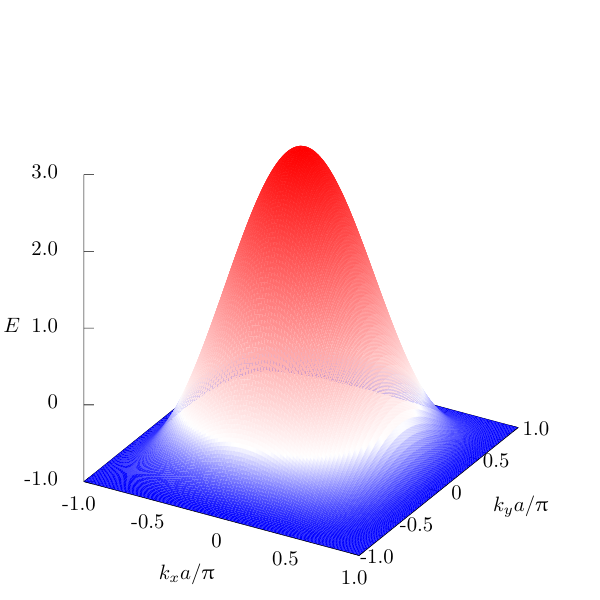}
\caption{}
\label{Fig_6}
\end{figure}
\enni \begin{figure}[h!]
\includegraphics[width=1.0\columnwidth]{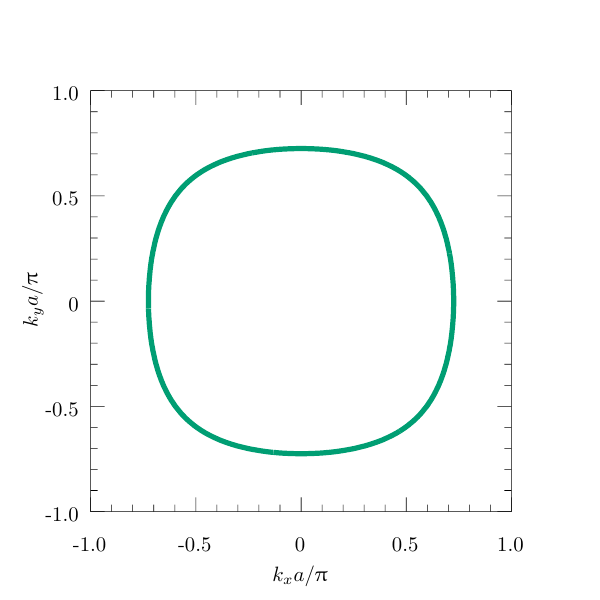}
\caption{}
\label{Fig_7}
\end{figure}

\clearpage
\pagebreak

\setcounter{equation}{0}

\widetext
\begin{center}
\textbf{Multiorbital singlet pairing and 
$d+d$ superconductivity \\
- Supplementary Information -
}
\end{center}

\makeatletter

\input{d_d_arxiv_SI}

\end{document}

%% file: d_d_arxiv_SI.tex
\section*{Supplementary Note 1: Pairing with spin structure - Lessons from superfluid \enhe}
\label{Sec:Appn_A}

The $p$-wave spin-triplet pairing in \enhe leads to a great variety of superfluid phases. These states are captured by a non-trivial matrix structure of the pairing in spin space. Here, we review some of the most important superfluid phases of \enhe paying particular attention to the form of the pairing matrix. 
In the main text (subsections A and B of the Results in the main text),
we draw analogies between superfluid \enhe and spin-singlet multiband SCs where the matrix structure is associated with local orbital DOF. These analogies can be made in spite of such obvious differences as the real-space symmetry which is continuous in \enhe and discrete in the SC cases and the normal-state bands which are typically more complex in SCs than in \enhe case.

Cooper pairing is the basic feature of all SCs. Electrons are bound in pairs to form a state which
saves ground-state energy.
 Even in the simplest BCS description of Cooper pairing, the spin-1/2 DOF of the two electrons play a crucial role. 
 The
 bound state has a net zero total spin. Paired electrons accordingly behave like bosons which 
 effectively undergo condensation into the superconducting ground state. 
 In turn, this makes it possible to understand the emergence of superconductivity as macroscopic breaking 
 of a $U(1)$ symmetry associated with the net condensate. This macroscopic breaking of symmetry underlies 
 most of the remarkable properties observed in experiments~\cite{de_Gennes, Vollhardt}. 
 The success of the basic picture of superconductivity is due in large part to the fundamental concepts 
 of spontaneous symmetry-breaking with little reliance on the detailed microscopic knowledge 
 of the interactions responsible for the Cooper instability.  

In addition to the net spin, Cooper pairs can also have net relative 
angular momentum (AM).

We restrict the discussion to homogeneous SC states. The symmetry of the Cooper pairs in that case is captured within a BdG formalism by the pairing function

\enni \begin{align}
\hat{\mathit{\Delta}}_{\alpha \beta}(\mathbf{k}) = \mathit{\Delta}(k) \hat{g}_{\text{Relative~AM} \alpha \beta}(\mathbf{\hat{k}}).  
\end{align} 

\enni $\mathit{\Delta}(k)$ is a pairing amplitude. 
The matrix elements of $\hat{g}_{\text{Relative~AM}}$ denote the weights of the tensor-product states of the local DOF. These weights are in general functions of the relative AM of the Cooper pair. In the simplest spin-1/2 singlet-pairing case, these matrix elements are the Clebsch-Gordan coefficients of the total spin $S=0$ of the Cooper pair.

In the absence of detailed microscopic models, symmetry provides a powerful tool in understanding the nature of the transition to the paired state as well as a host of the latter's experimental signatures. The case of superfluid \enhe is illustrative. The basic microscopic constituents are essentially featureless spin-1/2 atoms. Normal \enhe is a Fermi liquid which preserves rotational invariance in real space, spin space, and a global phase~\cite{Vollhardt}. Mathematically, this defines a group $G = SO(3)_{\mathbf{L}} \otimes SO(3)_{\mathbf{S}} \otimes U(1)$. Interactions which lead to Cooper pairing preserve this symmetry. In the weak-coupling regime, the pairing amplitude is dwarfed by an energy scale which characterizes the normal liquid~\cite{Anderson}. To a good approximation, the interactions and the pairing function can be projected onto sectors belonging to one of the irreducible representations of $G$~\cite{Anderson, Vollhardt}. Each representation is assigned a critical temperature $T_{\text{c}}$ determined by the eigenvalues of the linearized gap equation. The irreducible representation with the highest $T_{\text{c}}$ determines the nature of the paired state. 
Instead of going into the microscopic analyses
~\cite{Levin},
 we will focus on symmetry-based analyses for the superfluid \enhe. 
The
 $p$-wave spin-triplet channel with relative AM $L=1$ and total spin $S=1$ was singled-out as the most likely candidate~\cite{Anderson}. 

The only essential local DOF in \enhe are the spin-1/2 moments of the pairing atoms. 
In the physically-relevant triplet sector the order parameter can be written as~\cite{Sigrist}

\enni \begin{align}
\hat{g}_{\text{Relative~AM} \alpha \beta} = & \left( \mathbf{d}(\enht) \cdot \mathbf{\upsigma} \right) \text{i}\sigma_{2} \notag \\
= & \begin{pmatrix}
-d_{1}(\mathbf{\hat{k}}) + \text{i}d_{2}(\mathbf{\hat{k}}) & d_{3}(\mathbf{\hat{k}}) \\
d_{3}(\mathbf{\hat{k}}) & d_{1}(\mathbf{\hat{k}}) + \text{i}d_{2}(\mathbf{\hat{k}})
\end{pmatrix},
\label{Eq:Gnrl_3He}
\end{align} 

\enni where the $\mathbf{\upsigma}$ Pauli matrices act on the spin-1/2 tensor-product states. Odd-parity, spin-exchange symmetric pairing is restricted to the three complex components of $\mathbf{d}$, which transforms as a vector under spin rotations~\cite{Vollhardt}. The physical significance of this vector is important. For unitary states, where $\text{i} \mathbf{d} \times \mathbf{d}^{*}=0$, it can be shown~\cite{Leggett, Vollhardt} that the vector $\mathbf{d}$ is real within a $\mathbf{\hat{k}}$-dependent overall phase. In these cases, 
for any given point on the Fermi surface, the spin state of the pair is
an eigenstate of the total spin projected along a direction with zero
eigenvalue; the unit vector along this direction specifies $\mathbf{d}$.

Due to the separate $SO(3)$ symmetry of both relative AM and spin sectors, the order parameter has a large degeneracy. The superfluid phases of \enhe can be understood in great part via the breaking of the symmetry of $G$ in order to relieve the inherent degeneracy. The local spin DOF play a crucial role in the nature of the resulting phases, as we now proceed to review. 

\section*{Supplementary Note 2: Contrasting B and A phases}

In zero magnetic field and for most pressures, \enhe condenses directly into the superfluid B phase. At low temperatures it is the most stable state. The B phase came to be identified with a unitary, "Balian-Werthamer" (BW)~\cite{Balian} order-parameter. In it's simplest form, the BW vector order-parameter is

\enni \begin{align}
\mathbf{d}_{BW} = \mathbf{\hat{k}}.
\label{Eq:BW}
\end{align}

\enni The orientation of the spins and of the relative AM are locked together, reflecting the breaking of symmetry down to a group $G'=SO(3)_{\mathbf{L}+ \mathbf{S}}$ of simultaneous rotations. The simplest BW state then corresponds to 
a
 unique $J=0$ irreducible representation of $G'$. The gap, given by 

\enni \begin{align}
\hat{\mathit{\Delta}}(\mathbf{k}) \hat{\mathit{\Delta}}^{\dag}(\mathbf{k}) \sim \mathit{\Delta}^{2}(k) \sigma_{0}
\end{align}

\enni is finite everywhere along the FS for both spin-1/2 quasiparticles. The emergence of the uniform gap can formally be traced to the anti-commuting Pauli matrices. Therefore, it is clear that the local spin structure of this pairing plays a very important role. Due to the uniform gap, the BW order-parameter is the most favored configuration at weak-coupling~\cite{Balian}. 

In a restricted region of the pressure-temperature phase diagram, \enhe condenses directly into the A phase described by an unitary "Anderson-Brinkman-Morel" (ABM) order-parameter vector~\cite{Anderson, Vollhardt} 

\enni \begin{align}
\mathbf{d}_{\text{ABM}} =\mathbf{ \hat{d}} \left( \mathbf{\hat{k}} \cdot \mathbf{\hat{m}} + \text{i} \mathbf{\hat{k}} \cdot \mathbf{\hat{n}} \right).
\label{Eq:ABM}
\end{align}

\enni The system spontaneously selects preferred quantization axes for the spin and relative AM of the pair, respectively. Equivalently, the symmetry is spontaneously broken to a group $G'=U_{L_{z} -\phi} \times U_{S_{z}}$, where $\phi$ denotes a global phase rotation. This is a unitary order parameter and the vector $\mathbf{\hat{d}}$ thus determines a fixed direction in spin-space along which the pairs have a vanishing dipole-moment $\braket{\mathbf{S}}_{\text{Pair}}=0$. For example, when the vector $\mathbf{\hat{d}}$ lies in the $xy$ plane, the system exhibits equal-spin pairing in each of the independent sectors with total $S_{z} = \pm 1$ while pairing does not occur for opposite-spin partners. The relative AM likewise acquires a preferred direction which is determined by the  triad of mutually-perpendicular unit vectors $\mathbf{\hat{m}}, \mathbf{\hat{n}}$ and $\mathbf{\hat{l}}$. The latter determines the direction in $\mathbf{\hat{k}}$-space where the gap vanishes and thus where nodes occur. This phase also 
breaks
 time-reversal (TR) symmetries
~\cite{Balatsky}. Due to the appearance of nodes, the ABM order-parameter is not stable at weak-coupling where the BW is always preferred~\cite{Vollhardt, Anderson}. 
 The
 ABM pairing is 
 stabilized
 when feedback effects are taken into account~\cite{Vollhardt}. 
 
 The $P-T$ phase diagram reflects this, as the A phase only occurs in a restricted region and always gives way to the B phase at lower temperatures~\cite{Vollhardt}.  

The phases of \enhe in magnetic fields $\mathbf{H}$ are also instructive. The order parameters in these cases turn out to be non-unitary. These phases also manifest orientation effects due to preferred axes for the spins of the paired electrons. To illustrate, we focus on the A phase. Near $T_{\text{c}}$ and under an applied field, the A phase gives way to a non-unitary A$_{1}$ phase. Pairing occurs in only one of the sectors with total $S_{z} = \pm 1$, as a consequence of the lifting of the degeneracy of the spin-up and -down FSs. For lower temperatures, the non-unitary A$_{2}$ phase stabilizes. It is described by the complex vector 

\enni \begin{align}
\mathbf{d}_{A_{2}} = \left[ A \hat{\mathbf{d}} + \text{i} B \mathbf{\hat{e}} \right] \left( \mathbf{\hat{k}} \cdot \mathbf{\hat{m}} + \text{i} \mathbf{\hat{k}} \cdot \mathbf{\hat{n}} \right).
\end{align}

\enni Pairing now occurs in both of the sectors of total $S_{z} = \pm 1$. However, due to the split of the FS under applied field, the two coefficients $A (B) \sim \mathit{\Delta}_{\uparrow \uparrow} \pm \mathit{\Delta}_{\downarrow \downarrow}$ reflect different pairing amplitudes for the two sectors. Both A$_{1,2}$ phases are non-unitary pairing states where 
$\mathit{\Delta}^{\dag}(\mathbf{k}) \mathit{\Delta}(\mathbf{k})$ is not simply proportional to the identity matrix. In such cases, it can be shown that a pair at $\mathbf{k}$ acquires a finite spin dipole-moment~\cite{Leggett, Vollhardt}:

\enni \begin{align}
\braket{\mathbf{S}} = \left( \text{i} \mathbf{d}(\mathbf{\hat{k}}) \times \mathbf{d}^{*}(\mathbf{\hat{k}}) \right) \cdot \hat{\mathit{\Delta}}(\mathbf{k}) \hat{\mathit{\Delta}}^{\dag}(\mathbf{k}).
\end{align}

\enni In the simplest cases, this spin-dipole moment points along the applied field $\mathbf{H}$.  The vectors $\mathbf{\hat{d}}$ and $ \mathbf{\hat{e}}$ are perpendicular to the applied field, while the local moment is parallel to a mutually-perpendicular direction $\mathbf{\hat{f}}= \mathbf{\hat{d}} \times \mathbf{\hat{e}}$. 

A comparison of A and B phases of superfluid \enhe is important for our analysis of SC phases with non-trivial orbital local DOF. The B phase is the 
more
 symmetric of the two in the sense that it's residual-symmetry is still enhanced with respect to that of the A phase. The spin and relative AM DOF are locked to produce a maximal uniform gap. This phase is consequently always preferred at lower temperatures. The more constrained A phase, which further breaks the rotation and TR symmetries has nodes and is only stabilized
  at higher pressure and temperature. We believe that these salient properties are not restricted to superfluid \enhe but can also manifest in solid-state SC with non-trivial local DOF. Thus, we expect that pairing can naturally take advantage of additional structure due to local orbital DOF to induce the strongest and most uniform gap along the FS, as in \enhe -B.   

\section*{Supplementary Note 3: Landau-Ginzburg theory for two coexisting $B_{1g}$ channels}

\label{Sec:Appn_B}

Here we construct a generalized LG theory following the arguments of subsection A of the Results of the main text.
 
We consider pairing withing a two-orbital $xz, yz$ model. We restrict our discussion to intra-orbital pairing interactions

\enni \begin{align}
V_{\alpha \beta; \gamma \delta}(\mathbf{k},\mathbf{k}') = V(\mathbf{k},\mathbf{k}')\delta_{\alpha \delta} \delta_{\beta \gamma} \delta_{\alpha \beta}
\end{align}

\enni where the Greek indices denote the two orbitals. Such pairing interactions can originate from a $t-J_{1}-J_{2}$ model~\cite{Goswami, Yu, Nica_Yu}. For these types of interactions, inter-orbital pairing characterized by the off-diagonal $\tau_{1}$ matrix are suppressed. Furthermore, we consider that $s_{x^{2}y^{2}}$ and $d_{x^{2}-y^{2}}$ channels are equally-favored. This would correspond to $J_{1}\approx J_{2}$ in a $t-J_{1}-J_{2}$ model. As discussed in subsection A of the Results of the main text, we expect that orbital structure plays an important role near this point. Hence, we consider two, nearly-degenerate $s_{x^{2}y^{2}}\tau_{3}$ and $d_{x^{2}-y^{2}} \tau_{0}$ order parameters. Each of these transforms as the single-component $B_{1g}$ irreducible representation of $D_{4h}$. In spite of this, it is important to distinguish between these two channels, since they result in different BdG spectra in the most general case. Indeed, in the two-band model for the Fe-based SC's considered previously, $d_{x^{2}-y^{2}} \tau_{0}$ commutes with the TB Hamiltonian in Eq.~7 of the main text, while $s_{x^{2}y^{2}}\tau_{3}$ does not.

We denote the order-parameters associated with $s_{x^{2}y^{2}}\tau_{3}$ and $d_{x^{2}-y^{2}} \tau_{0}$ by $\mathit{\Delta}_{1}$ and $\mathit{\Delta}_{2}$, respectively. The general LG free-energy for these two candidates is 

\begin{widetext}

\enni \begin{align}
F_{LG} = & F^{(2)} + F^{(4)} \\
F^{(2)} = & \alpha_{1} |\mathit{\Delta}_{1}|^{2} + \alpha_{2} |\mathit{\Delta}_{1}|^{2} + \alpha_{3} \left( \mathit{\Delta}^{*}_{1}\mathit{\Delta}_{2} + \mathit{\Delta}_{1}\mathit{\Delta}^{*}_{2}\right)  \\
F^{(4)} = & \beta_{1}|\mathit{\Delta}_{1}|^{4} + \beta_{2}|\mathit{\Delta}_{2}|^{2} + 
\beta_{3} |\mathit{\Delta}_{1}|^{2}|\mathit{\Delta}_{2}|^{2} + \beta_{4}\left[(\mathit{\Delta}^{*}_{1} \mathit{\Delta}_{2})^{2} + (\mathit{\Delta}_{1} \mathit{\Delta}^{*}_{2})^{2}  \right] + \beta_{5} |\mathit{\Delta}_{1}|^{2} \left( \mathit{\Delta}^{*}_{1}\mathit{\Delta}_{2} + \mathit{\Delta}_{1}\mathit{\Delta}^{*}_{2}\right) \notag \\ 
+ & \beta_{6} |\mathit{\Delta}_{2}|^{2} \left( \mathit{\Delta}^{*}_{1}\mathit{\Delta}_{2} + \mathit{\Delta}_{1}\mathit{\Delta}^{*}_{2}\right)
\end{align}

\end{widetext}

\enni Terms involving $\alpha_{1,2}$ and $\beta_{1-4}$ are also allowed in the case of almost-degenerate but orbital-trivial $s$ and $d$ order parameters~\cite{Lee_Zhang_Wu}. Couplings involving $\alpha_{3}, \beta_{5}$, and $\beta_{6}$ are made possible by the orbital structure. We assume that $\alpha_{3}$ is independent of temperature. 

In principle, the phase diagram of this model can be determined by minimizing with respect to the two amplitudes and the relative phase of the two components for arbitrary values of the coefficients. This is beyond the scope of this section which is to show how alternatives to a TR-broken state can emerge.
To make progress, we introduce normalized pairing coefficients as

\enni \begin{align}
\mathit{\Delta}_{1} = & \cos(\theta) e^{i\phi_{1}} |\mathit{\Delta}| \\
\mathit{\Delta}_{2} = & \sin(\theta) e^{i\phi_{2}} |\mathit{\Delta}|,
\end{align} 

\enni with $\theta \in [0, \uppi/2]$ as a convenient way of parameterizing the two pairing channels
. The LG free-energy can be written as 

\begin{widetext}
\enni \begin{align}
F_{LG} = & \tilde{\alpha} |\mathit{\Delta}|^{2} + \tilde{\beta} |\mathit{\Delta}|^{4} \\
\tilde{\alpha} = & \alpha_{1}\cos^{2}(\theta) + \alpha_{2}\sin^{2}(\theta) + 2 \alpha_{3} \cos(\phi) \cos(\theta) \sin(\theta) \\
\label{Eq:Eff_alph}
\tilde{\beta} = & \beta_{1}\cos^{4}(\theta) + \beta_{2}\sin^{4}(\theta) + \beta_{3}\cos^{2}(\theta)\sin^{2}(\theta) + 2 \beta_{4} \cos(2\phi) \cos^{2}(\theta)\sin^{2}(\theta) 
+ 2\beta_{5} \cos(\phi) \cos^{3}(\theta) \sin(\theta) \notag \\
+ & 2\beta_{6} \cos(\phi) \cos(\theta) \sin^{3}(\theta),
\end{align}
\end{widetext}

\enni where $\phi=\phi_{1}-\phi_{2}$ is the relative phase of the two components. 

In any SC phase, we require that $\tilde{\alpha} <0, \tilde{\beta}>0$. We assume that near the superconducting transition, quartic terms can essentially be neglected. The minimum of $F_{LG}$ is then determined by the minimum of $\tilde{\alpha}$. For given $\alpha_{1-3}$, we find the extrema of $\tilde{\alpha}$  as a function of $\theta, \phi$ for coexisting channels with $\theta \neq 0, \uppi/2$.  We obtain:

\enni \begin{align}
\tan(2\theta) = & \frac{2\alpha_{3} \cos(\phi)}{\alpha_{1} - \alpha_{2}} \\
-\alpha_{3} \sin(\phi) \sin(2\theta)= & 0. 
\end{align}

\enni The second condition implies $\phi = 0$ or $\uppi$. Using double-angle formulas,  the first condition is

\enni \begin{align}
\tan(\theta) = \frac{\alpha_{1} - \alpha_{2} \pm \sqrt{(\alpha_{1} - \alpha_{2})^{2} + 4(\alpha_{3} \cos(\phi))^{2}}}{-2\alpha_{3} \cos(\phi)}.
\end{align} 

For simplicity we consider $\alpha_{3}>0$. To see what the extrema determined above imply, we consider two limiting cases. First, we consider almost degenerate channels with $|(\alpha_{1} - \alpha_{2})/2 \alpha_{3}| \ll 1$. The two extrema correspond to leading order to 

\enni \begin{align}
\theta \approx \frac{\uppi}{4} \pm \frac{\alpha_{1} - \alpha_{2}}{4 \alpha_{3}} +  O\left[ \left(\frac{\alpha_{1} - \alpha_{2}}{2 \alpha_{3}} \right)^{2} \right]
\end{align}

\enni and 

\enni \begin{align}
\tilde{\alpha} = & \mp \alpha_{3}\left\{1 \mp \frac{\alpha_{1}+\alpha_{2}}{2\alpha_{3}} +  O\left[ \left(\frac{\alpha_{1} - \alpha_{2}}{2 \alpha_{3}} \right)^{2} \right] \right\}
\end{align}

\enni for $\phi=\uppi$ and 0, respectively.
It is straightforward to show that the solution corresponding to $\phi=\uppi$ is always smaller than either $\alpha_{1}$ or $\alpha_{2}$ which correspond to a pairing in either $s_{x^{2}y^{2}}\tau_{3}$ or $d_{x^{2}-y^{2}}\tau_{0}$ channels below $T_{\text{c}}$. This implies that the bilinear terms in the LG theory always prefer a coexisting solution immediately  below $T_{\text{c}}$.   
  
The other limiting cases occur for   
$|(\alpha_{1} - \alpha_{2})/2 \alpha_{3}| \gg 1$. In this case, $\theta$ is close to either $0$ or $\uppi/2$. Below $T_{\text{c}}$ one channel will dominate with a very small admixture of the other.  

In general, we expect that no additional transitions occur below $T_{\text{c}}$ although this can occur in some instances as illustrated by the analogous case of \enhe -A$_{1}$ and \enhe -A$_{2}$. 

These arguments illustrate that coexisting $s\tau_{3}$ and $d\tau_{0}$ channels are in general favored when pairing interactions are similar in the two cases. 

\section*{Supplementary Note 4: $s\tau_{3}$ pairing in the band basis}

\label{Sec:stau3_bnd}

For a general effective two-orbital $d_{xz,yz}$ model, the factors entering Eq.~7 of the main text are~\cite{Raghu, Nica_Yu}

\enni \begin{align}
\xi_{0}(\mathbf{k}) = & -(t_{1} + t_{2})( \cos(k_{x}a) + \cos(k_{y}a) ) \notag \\
- & 4t_{3}\cos(k_{x}a) \cos(k_{y}a) \notag \\
\xi_{1}(\mathbf{k}) = & -4t_{3} \sin(k_{x}a) \sin(k_{y}a) \notag \\
\xi_{3}(\mathbf{k}) = & -(t_{1} - t_{2})( \cos(k_{x}a) - \cos(k_{y}a) ),
\end{align}

\enni where $t_{1-4}$ are TB parameters~\cite{Raghu}.
Note that $\xi_{0}, \xi_{1}$, and $\xi_{3}$ transform as $A_{1g}$, $B_{2g}$, and $B_{1g}$ representations of $D_{4g}$, respectively.

To diagonalize the TB part of Eq.~7 of the main text, we use the hermitian matrix 

\enni \begin{align}
V(\mathbf{k})
= & \frac{1}{\sqrt{2}}
\begin{pmatrix}
-\sqrt{1-c(\mathbf{k})} & A(\mathbf{k}) \sqrt{1+c(\mathbf{k})} \\
\\
A(\mathbf{k}) \sqrt{1+c(\mathbf{k})} & \sqrt{1-c(\mathbf{k})}
\end{pmatrix}, 
\end{align}

\enni where 

\enni \begin{align}
c(\mathbf{k}) = & \frac{\xi_{3}(\mathbf{k})}{\sqrt{\xi^{2}_{1}(\mathbf{k}) + \xi^{2}_{3}(\mathbf{k})}}, 
\\
A(\mathbf{k}) = & \text{sgn}\left( \xi_{1}(\mathbf{k}) \right).
\end{align}

\enni The factors $1\pm c(\mathbf{k}) \ge 0$ since $-1 \le c \le 1$. Note that $V(\mathbf{k})$ is parity-even due to $\xi_{1,3}$ also being even. For convenience, we also define the factors 

\enni \begin{align}
s(\mathbf{k}) = & \frac{\xi_{1}(\mathbf{k})}{\sqrt{\xi^{2}_{1}(\mathbf{k}) + \xi^{2}_{3}(\mathbf{k})}},
\end{align}

\enni which obey $c^{2}+s^{2}=1$ for all $\mathbf{k}$. It is straightforward to show that $V(\mathbf{k}) \hat{H}_{\text{TB}} V^{\dag}(\mathbf{k})$ transforms the TB Hamiltonian from the orbital to the band basis. 

The form of $s\tau_{3}$ pairing in the band basis (Eq.~10 in the main text) is obtained via the transformation 

\enni \begin{align}
\hat{H}_{\text{Pair, Band}} =V(\mathbf{k}) \hat{H}_{\text{Pair, Orbital}} V^{T}(-\mathbf{k}). 
\end{align}

\enni Carrying out this transformation, we obtain the factors $\mathit{\Delta}_{1,3}(\mathbf{k})$ in Eq.~10 in the main text

\enni \begin{align}
\mathit{\Delta}_{3}(\mathbf{k}) = & -
s_{x^{2}y^{2}}(\mathbf{k})
\frac{\xi_{3}(\mathbf{k})}{\sqrt{\xi^{2}_{1}(\mathbf{k}) + \xi^{2}_{3}(\mathbf{k})}} \\
\mathit{\Delta}_{1}(\mathbf{k}) = & - s_{x^{2}y^{2}}(\mathbf{k})
\frac{\xi_{1}(\mathbf{k})}{\sqrt{\xi^{2}_{1}(\mathbf{k}) + \xi^{2}_{3}(\mathbf{k})}}.
\end{align}

\enni These expressions suffer from a sign ambiguity when either $\xi_{1/3} \rightarrow 0$. However, since the TB and pairing parts commute, the gap is proportional to  $\text{Tr}\hat{\mathit{\Delta}}^{*} \hat{\mathit{\Delta}}$, and the overall sign of the pairing has no effect on the physical BdG spectrum. Since $\xi_{3}, \xi_{1}$ transform as $B_{1g}$ and $B_{2g}$ representations, respectively, so do the two intra- and inter-band components, as discussed in the main text. 
We note that terms proportional to $\alpha_{0}$ do not occur due to the absence of corresponding $\tau_{0}$ terms in the orbital basis. Similarly, terms proportional to $\alpha_{2}$ are anti-symmetric under band exchange and are excluded due to the absence of anti-symmetric terms in the orbital basis.

To see how the matrices $\alpha_{1,3}$ in band space transform under the point group, we apply an inverse transformation from band to orbital space. In the orbital basis, the transformation properties of the matrices are well-defined. We first consider the matrix $\alpha_{3}$:

\enni \begin{align}
V^{T}(\mathbf{k}) \alpha_{3} V(-\mathbf{k}) 
= & -\frac{\xi_{3}}{\sqrt{\xi^{2}_{1}+\xi^{2}_{3}}} \tau_{3} -\frac{\xi_{1}}{\sqrt{\xi^{2}_{1}+\xi^{2}_{3}}} \tau_{1}.
\end{align} 

\enni The first component transforms as $B_{1g} \times B_{1g} = A_{1g}$. For the second, we use the fact that $\tau_{1}$ transforms as $B_{2g}$~\cite{Wan_Wang}. 
Together with the form factor it transforms as $B_{2g} \times B_{2g} = A_{1g}$. 
Hence, $\alpha_{3}$ is equivalent to an $A_{1g}$ representation. We next consider the $\alpha_{1}$ matrix:

\enni \begin{align}
V^{T}(\mathbf{k}) \alpha_{1} V(-\mathbf{k}) 
= & -\frac{\xi_{1}}{\sqrt{
\xi^{2}_{1} 
+ \xi^{2}_{3}}} \tau_{3} + \frac{\xi_{3}}{\sqrt{\xi^{2}_{1} + \xi^{2}_{3}}}\tau_{1} 
\end{align}

\enni The first term transforms as $B_{2g} \times B_{1g} = A_{2g} $ and the second term transforms as $B_{1g} \times B_{2g}=A_{2g}$. Therefore, the matrix $\alpha_{1}$ is equivalent to an $A_{2g}$ representation.

The same results can also be obtained by demanding that $\xi_{3} \alpha_{3}$ and $\xi_{1} \alpha_{1}$ each belong to $B_{1g}$ since the pairing in the band and orbital basis are equivalent.

\section*{Supplementary note 5:Pairing of damped quasiparticles}
\label{Sec:Appn_I}

In subsection A of the Results of the main text
, we mentioned that there can be two cases with $s\tau_{3}$
at the level of the BdG equations. While the single-particle excitations are always gapped on the FS, depending on the details of the bandstructure they can: I) remain gapped away  from the FS (as in the case of the alkaline iron selenides) or II) lead to the emergence of nodes away from the FS. We now consider what happens to the type-II
case when we take into account the damping of the quasiparticles away from the FS being nonzero, as must be the case due to underlying correlations. Such damping suppresses pairing, thereby diminishing the experimental signatures of such away-from-FS nodes. This is to be contrasted to the case of nodes on the FS, where the assumption of long-lived quasiparticles is robust.

Here, we provide proof-of-concept calculations for the suppression of pairing for quasiparticles with finite lifetimes. We consider a canonical, weak-coupling BCS paradigm with additional \emph{ad hoc} momentum and frequency-independent lifetimes. We believe that this simple approach illustrates our statement although quasiparticle damping is model-dependent in general. 

We consider the $T=0$ time-ordered Green's functions for the standard BCS case~\cite{Schrieffer} with the inclusion of a general momentum- and frequency-independent damping term via the replacement

\enni \begin{align}
\epsilon_{\mathbf{k}} \rightarrow \epsilon_{\mathbf{k}} - \text{i} \text{sgn}(\omega) \mathit{\mathit{\Gamma}},
\end{align} 

\enni where $\epsilon_{\mathbf{k}}$ is the usual dispersion and the second term with $\mathit{\mathit{\Gamma}} >0$ independent of momentum and frequency introduces a finite damping and respects the analytic structure of time-ordered functions. We assume static pairing terms $\mathit{\Delta}_{\mathbf{k}}$, in order to illustrate finite-lifetime effects relative to the canonical case where $\mathit{\Gamma} \rightarrow 0$. We impose a standard BCS self-consistency condition at $T=0$

\enni \begin{align}
\mathit{\Delta}_{\mathbf{k}} = & i \lim_{t \rightarrow 0^{-}} \frac{1}{2} \sum_{\mathbf{q}} V_{\mathbf{k-q}}\text{Tr} \left[ \tau_{1} \bm{G}(t, \mathbf{q}) \right],   
\end{align} 

\enni where $\bm{G}$ is the matrix Green's function in the Nambu-Gor'kov formalism~\cite{Schrieffer}. We re-cast the resulting gap equation at zero-temperature in the form

\enni \begin{align}
1 -  \frac{\rho_{0} V_{0}}{2} I\left( \frac{\omega_{D}}{|\mathit{\Delta}|}, \frac{\mathit{\Gamma}}{|\mathit{\Delta}|} \right) =0
\label{Eq:Stnd_frm}
\end{align}

\enni where $\rho_0$ is the DOS, $V_{0}$ represents the momentum-independent pairing interactions within a cutoff $\pm \omega_{D}$ near the Fermi level, respectively, and where we assumed a featureless $s$-wave pairing with $|\mathit{\Delta}|_{\mathbf{k}} = |\mathit{\Delta}|$. $I$ is defined by the dimensionsless integral 

\enni \begin{align}
I\left( \frac{\omega_{D}}{|\mathit{\Delta}|}, \frac{\mathit{\Gamma}}{|\mathit{\Delta}|} \right) 
= &  
\int^{\frac{\omega_{D}}{|\mathit{\Delta}|}}_{-\frac{\omega_{D}}{|\mathit{\Delta}|}} \frac{d x}{ \sqrt{\left( x - i \frac{\mathit{\Gamma}}{|\mathit{\Delta}|} \right)^{2} + 1}}
\label{Eq:Dmsn_int}
\end{align}

In the usual BCS case with $\mathit{\Gamma} =0$, $I$ diverges logarithmically as $|\mathit{\Delta}| \rightarrow 0^{+}$, ensuring that solutions to the gap equation can always be found for arbitrarily weak interactions ($\rho_{0} V_{0} \rightarrow  0$). 

For general damping $\mathit{\Gamma} \neq 0 $, we consider what happens when the gap equation is linearized. It follows from Eq.~\ref{Eq:Dmsn_int} that the finite damping cuts off the logarithmic divergence in the usual BCS case since 

\enni \begin{align}
\lim_{|\mathit{\Delta}| \rightarrow 0^{+}} I = & \ln \left[ \frac{\omega^{2}_{D} + \mathit{\Gamma}^{2}}{\mathit{\Gamma}^{2}} \right].
\end{align}

\enni This can be checked by carrying out the integral in the complex plane with care to avoid the branch cuts. Hence, the presence of strong damping imposes a nonzero lower bound on the pairing interaction strength such that zero-temperature solutions of the gap equations $|\mathit{\Delta}|$ are not possible below this threshold. This shows that the pairing of strongly damped quasiparticles is always suppressed relative to the case without damping. Although our conclusions are based on a very simple model, we expect that similar effects are present in more realistic calculations.



%% file: d_d_arxiv.bbl
\section*{References}

\begin{thebibliography}{10}
\expandafter\ifx\csname url\endcsname\relax
  \def\url#1{\texttt{#1}}\fi
\expandafter\ifx\csname urlprefix\endcsname\relax\def\urlprefix{URL }\fi
\providecommand{\bibinfo}[2]{#2}
\providecommand{\eprint}[2][]{\url{#2}}

\bibitem{Lee_2006}
\bibinfo{author}{Lee, P.~A.}, \bibinfo{author}{Nagaosa, N.} \&
  \bibinfo{author}{Wen, X.-G.}
\newblock \bibinfo{title}{Doping a {M}ott insulator: physics of
  high-temperature superconductivity}.
\newblock \emph{\bibinfo{journal}{Rev. Mod. Phys.}}
  \textbf{\bibinfo{volume}{78}}, \bibinfo{pages}{17--85}
  (\bibinfo{year}{2006}).

\bibitem{Scalapino_2012}
\bibinfo{author}{Scalapino, D.~J.}
\newblock \bibinfo{title}{{A common thread: the pairing interaction for
  unconventional superconductors}}.
\newblock \emph{\bibinfo{journal}{Rev. Mod. Phys.}}
  \textbf{\bibinfo{volume}{84}}, \bibinfo{pages}{1383--1417}
  (\bibinfo{year}{2012}).

\bibitem{Agterberg_2017}
\bibinfo{author}{Agterberg, D.~F.}, \bibinfo{author}{Brydon, P. M.~R.} \&
  \bibinfo{author}{Timm, C.}
\newblock \bibinfo{title}{{Bogoliubov Fermi surfaces in superconductors with
  broken time-reversal symmetry}}.
\newblock \emph{\bibinfo{journal}{Phys. Rev. Lett.}}
  \textbf{\bibinfo{volume}{118}}, \bibinfo{pages}{127001}
  (\bibinfo{year}{2017}).

\bibitem{Ramires_2018}
\bibinfo{author}{Ramires, A.}, \bibinfo{author}{Agterberg, D.~F.} \&
  \bibinfo{author}{Sigrist, M.}
\newblock \bibinfo{title}{Tailoring ${T}_{\text{c}}$ by symmetry principles: the
  concept of superconducting fitness}.
\newblock \emph{\bibinfo{journal}{Phys. Rev. B}} \textbf{\bibinfo{volume}{98}},
  \bibinfo{pages}{024501} (\bibinfo{year}{2018}).

\bibitem{Hosono}
\bibinfo{author}{Kamihara, Y.}, \bibinfo{author}{Watanabe, T.},
  \bibinfo{author}{Hirano, M.} \& \bibinfo{author}{Hosono, H.}
\newblock \bibinfo{title}{{Iron-based layered superconductor
  {L}a[{O}$_{1-x}${F}$_x$]{F}e{A}s ($x = 0.05 - 0.12$) with $T_{\text c} = 26$
  {K}}}.
\newblock \emph{\bibinfo{journal}{J. Am. Chem. Soc.}}
  \textbf{\bibinfo{volume}{130}}, \bibinfo{pages}{3296--3297}
  (\bibinfo{year}{2008}).

\bibitem{Johnston2011}
\bibinfo{author}{Johnston, D.~C.}
\newblock \bibinfo{title}{{The puzzle of high temperature superconductivity in
  layered iron pnictides and chalcogenides}}.
\newblock \emph{\bibinfo{journal}{Adv. Phys.}} \textbf{\bibinfo{volume}{59}},
  \bibinfo{pages}{803--1061} (\bibinfo{year}{2010}).

\bibitem{Dai_RMP2015}
\bibinfo{author}{Dai, P.}
\newblock \bibinfo{title}{Antiferromagnetic order and spin dynamics in
  iron-based superconductors}.
\newblock \emph{\bibinfo{journal}{Rev. Mod. Phys.}}
  \textbf{\bibinfo{volume}{87}}, \bibinfo{pages}{855--896}
  (\bibinfo{year}{2015}).

\bibitem{Yi2017}
\bibinfo{author}{Yi, M.}, \bibinfo{author}{Zhang, Y.}, \bibinfo{author}{Shen,
  Z.-X.} \& \bibinfo{author}{Lu, D.}
\newblock \bibinfo{title}{Role of the orbital degree of freedom in iron-based
  superconductors}.
\newblock \emph{\bibinfo{journal}{npj Quantum Materials}}
  \textbf{\bibinfo{volume}{2}}, \bibinfo{pages}{57} (\bibinfo{year}{2017}).

\bibitem{Si2016}
\bibinfo{author}{Si, Q.}, \bibinfo{author}{Yu, R.} \&
  \bibinfo{author}{Abrahams, E.}
\newblock \bibinfo{title}{{High-temperature superconductivity in iron pnictides
  and chalcogenides}}.
\newblock \emph{\bibinfo{journal}{Nat. Rev. Mater.}}
  \textbf{\bibinfo{volume}{1}}, \bibinfo{pages}{16017} (\bibinfo{year}{2016}).

\bibitem{Wang_Lee2011}
\bibinfo{author}{Wang, F.} \& \bibinfo{author}{Lee, D.-H.}
\newblock \bibinfo{title}{{The electron-pairing mechanism of iron-based
  superconductors}}.
\newblock \emph{\bibinfo{journal}{Science}} \textbf{\bibinfo{volume}{332}},
  \bibinfo{pages}{200--204} (\bibinfo{year}{2011}).

\bibitem{Hirschfeld2016}
\bibinfo{author}{Hirschfeld, P.~J.}
\newblock \bibinfo{title}{{Using gap symmetry and structure to reveal the
  pairing mechanism in {F}e-based superconductors}}.
\newblock \emph{\bibinfo{journal}{C. R. Physique}}
  \textbf{\bibinfo{volume}{17}}, \bibinfo{pages}{197--231}
  (\bibinfo{year}{2016}).

\bibitem{Yi_prl2013}
\bibinfo{author}{Yi, M.} \emph{et~al.}
\newblock \bibinfo{title}{{Observation of temperature-induced crossover to an
  orbital-selective {M}ott phase in
  {${\mathrm{A}}_{x}{\mathrm{Fe}}_{2\mathrm{\text{\ensuremath{-}}}y}{\mathrm{Se}}_{2}$
  ($A\mathbf{=}\mathrm{K}$, Rb)} superconductors}}.
\newblock \emph{\bibinfo{journal}{Phys. Rev. Lett.}}
  \textbf{\bibinfo{volume}{110}}, \bibinfo{pages}{067003}
  (\bibinfo{year}{2013}).

\bibitem{Yu_prl2013}
\bibinfo{author}{Yu, R.} \& \bibinfo{author}{Si, Q.}
\newblock \bibinfo{title}{{Orbital-selective {M}ott phase in multiorbital
  models for alkaline iron selenides
  {${\mathbf{K}}_{1\ensuremath{-}x}{\mathrm{Fe}}_{2\ensuremath{-}y}{\mathrm{Se}}_{2}$}}}.
\newblock \emph{\bibinfo{journal}{Phys. Rev. Lett.}}
  \textbf{\bibinfo{volume}{110}}, \bibinfo{pages}{146402}
  (\bibinfo{year}{2013}).

\bibitem{Yu_Zhu_Si}
\bibinfo{author}{Yu, R.}, \bibinfo{author}{Zhu, J.-X.} \& \bibinfo{author}{Si,
  Q.}
\newblock \bibinfo{title}{{Orbital-selective superconductivity, gap anisotropy,
  and spin resonance excitations in a multiorbital $t$-${J}_{1}$-${J}_{2}$
  model for iron pnictides}}.
\newblock \emph{\bibinfo{journal}{Phys. Rev. B}} \textbf{\bibinfo{volume}{89}},
  \bibinfo{pages}{024509} (\bibinfo{year}{2014}).

\bibitem{Yin_Haule_Kotliar}
\bibinfo{author}{Yin, Z.~P.}, \bibinfo{author}{Haule, K.} \&
  \bibinfo{author}{Kotliar, G.}
\newblock \bibinfo{title}{Spin dynamics and orbital-antiphase pairing symmetry
  in iron-based superconductors}.
\newblock \emph{\bibinfo{journal}{Nat. Phys.}} \textbf{\bibinfo{volume}{10}},
  \bibinfo{pages}{845--850} (\bibinfo{year}{2014}).

\bibitem{Coleman}
\bibinfo{author}{Ong, T.}, \bibinfo{author}{Coleman, P.} \&
  \bibinfo{author}{Schmalian, J.}
\newblock \bibinfo{title}{{Concealed d-wave pairs in the s{\textpm} condensate
  of iron-based superconductors}}.
\newblock \emph{\bibinfo{journal}{Proc. Nat. Acad. Sci.}}
  \textbf{\bibinfo{volume}{113}}, \bibinfo{pages}{5486--5491}
  (\bibinfo{year}{2016}).

\bibitem{Nica_Yu}
\bibinfo{author}{Nica, E.~M.}, \bibinfo{author}{Yu, R.} \& \bibinfo{author}{Si,
  Q.}
\newblock \bibinfo{title}{Orbital-selective pairing and superconductivity in
  iron selenides}.
\newblock \emph{\bibinfo{journal}{npj Quantum Materials}}
  \textbf{\bibinfo{volume}{2}}, \bibinfo{pages}{24} (\bibinfo{year}{2017}).
\newblock \eprint{arXiv:1505.04170}.

\bibitem{Kreisel2017}
\bibinfo{author}{Kreisel, A.} \emph{et~al.}
\newblock \bibinfo{title}{Orbital selective pairing and gap structures of
  iron-based superconductors}.
\newblock \emph{\bibinfo{journal}{Phys. Rev. B}} \textbf{\bibinfo{volume}{95}},
  \bibinfo{pages}{174504} (\bibinfo{year}{2017}).

\bibitem{HY_Hu2019}
\bibinfo{author}{Hu, H.}, \bibinfo{author}{Yu, R.}, \bibinfo{author}{Nica,
  E.~M.}, \bibinfo{author}{Zhu, J.-X.} \& \bibinfo{author}{Si, Q.}
\newblock \bibinfo{title}{Orbital-selective superconductivity in the nematic
  phase of {F}e{S}e}.
\newblock \emph{\bibinfo{journal}{Phys. Rev. B}} \textbf{\bibinfo{volume}{98}},
  \bibinfo{pages}{220503} (\bibinfo{year}{2018}).

\bibitem{Hu_PRR_2020}
\bibinfo{author}{Hu, L.-H.}, \bibinfo{author}{Johnson, P.~D.} \&
  \bibinfo{author}{Wu, C.}
\newblock \bibinfo{title}{Pairing symmetry and topological surface state in
  iron-chalcogenide superconductors}.
\newblock \emph{\bibinfo{journal}{Phys. Rev. Res.}}
  \textbf{\bibinfo{volume}{2}}, \bibinfo{pages}{022021} (\bibinfo{year}{2020}).

\bibitem{C_Zhang2013}
\bibinfo{author}{Zhang, C.} \emph{et~al.}
\newblock \bibinfo{title}{Measurement of a double neutron-spin resonance and an
  anisotropic energy gap for underdoped superconducting
  {N}a{F}e$_{0.985}${C}o$_{0.015}${A}s using inelastic neutron scattering}.
\newblock \emph{\bibinfo{journal}{Phys. Rev. Lett.}}
  \textbf{\bibinfo{volume}{111}}, \bibinfo{pages}{207002}
  (\bibinfo{year}{2013}).

\bibitem{Sprau}
\bibinfo{author}{Sprau, P.~O.} \emph{et~al.}
\newblock \bibinfo{title}{{Discovery of orbital-selective Cooper pairing in
  FeSe}}.
\newblock \emph{\bibinfo{journal}{Science}} \textbf{\bibinfo{volume}{357}},
  \bibinfo{pages}{75--80} (\bibinfo{year}{2017}).

\bibitem{Shibauchi_arxiv_2020}
\bibinfo{author}{Shibauchi, T.}, \bibinfo{author}{Tetsuo, H.} \&
  \bibinfo{author}{Matsuda, Y.}
\newblock \bibinfo{title}{Exotic superconducting states in {F}e{S}e-based
  materials. {P}reprint at
  \href{https://arxiv.org/abs/2005.07315}{https://arxiv.org/abs/2005.07315}}
  (\bibinfo{year}{2020}).

\bibitem{Smidman}
\bibinfo{author}{Smidman, M.} \emph{et~al.}
\newblock \bibinfo{title}{{Interplay between unconventional superconductivity
  and heavy-fermion quantum criticality: {C}e{C}u$_{2}${S}i$_{2}$ versus
  {Y}b{R}h$_{2}${S}i$_{2}$}}.
\newblock \emph{\bibinfo{journal}{Philos. Mag.}} \textbf{\bibinfo{volume}{98}},
  \bibinfo{pages}{2930--2963} (\bibinfo{year}{2018}).

\bibitem{Park}
\bibinfo{author}{Park, J.~T.} \emph{et~al.}
\newblock \bibinfo{title}{{Magnetic resonant mode in the low-energy
  spin-excitation spectrum of superconducting {R}b$_{2}${F}e$_{4}${S}e$_{5}$
  single crystals}}.
\newblock \emph{\bibinfo{journal}{Phys. Rev. Lett.}}
  \textbf{\bibinfo{volume}{107}}, \bibinfo{pages}{177005}
  (\bibinfo{year}{2011}).

\bibitem{Friemel}
\bibinfo{author}{Friemel, G.} \emph{et~al.}
\newblock \bibinfo{title}{{Reciprocal-space structure and dispersion of the
  magnetic resonant mode in the superconducting phase of
  {R}b${}_{x}${F}e${}_{2\ensuremath{-}y}${S}e${}_{2}$ single crystals}}.
\newblock \emph{\bibinfo{journal}{Phys. Rev. B}} \textbf{\bibinfo{volume}{85}},
  \bibinfo{pages}{140511} (\bibinfo{year}{2012}).

\bibitem{Eschrig}
\bibinfo{author}{Eschrig, M.}
\newblock \bibinfo{title}{{The effect of collective spin-1 excitations on
  electronic spectra in high-T$_\text{c}$ superconductors}}.
\newblock \emph{\bibinfo{journal}{Adv. Phys.}} \textbf{\bibinfo{volume}{55}},
  \bibinfo{pages}{47--183} (\bibinfo{year}{2006}).

\bibitem{Stockert_Nat_Phys_2011}
\bibinfo{author}{Stockert, O.} \emph{et~al.}
\newblock \bibinfo{title}{{{Magnetically driven superconductivity in
  CeCu$_2$Si$_2$}}}.
\newblock \emph{\bibinfo{journal}{Nat. Phys.}} \textbf{\bibinfo{volume}{7}},
  \bibinfo{pages}{119--124} (\bibinfo{year}{2011}).

\bibitem{Maier_PRB_2011}
\bibinfo{author}{Maier, T.~A.}, \bibinfo{author}{Graser, S.},
  \bibinfo{author}{Hirschfeld, P.~J.} \& \bibinfo{author}{Scalapino, D.~J.}
\newblock \bibinfo{title}{{$d$-wave pairing from spin fluctuations in the
  ${\mathrm{K}}_{x}$Fe${}_{2\ensuremath{-}y}$Se${}_{2}$ superconductors}}.
\newblock \emph{\bibinfo{journal}{Phys. Rev. B}} \textbf{\bibinfo{volume}{83}},
  \bibinfo{pages}{100515} (\bibinfo{year}{2011}).

\bibitem{Mou}
\bibinfo{author}{Mou, D.} \emph{et~al.}
\newblock \bibinfo{title}{{Distinct Fermi surface topology and nodeless
  superconducting gap in a ({T}l$_{0.58}${R}b$_{0.42}$){F}e$_{1.72}${S}e$_{2}$
  superconductor}}.
\newblock \emph{\bibinfo{journal}{Phys. Rev. Lett.}}
  \textbf{\bibinfo{volume}{106}}, \bibinfo{pages}{107001}
  (\bibinfo{year}{2011}).

\bibitem{Wang_2011}
\bibinfo{author}{{X.-P. Wang and T. Qian and P. Richard and P. Zhang and J.
  Dong and H.-D. Wang and C.-H. Dong and M.-H. Fang and H. Ding}}.
\newblock \bibinfo{title}{{Strong nodeless pairing on separate electron Fermi
  surface sheets in ({T}l, {K}){F}e$_{1.78}${S}e$_2$ probed by {ARPES}}}.
\newblock \emph{\bibinfo{journal}{Europhy. Lett.}}
  \textbf{\bibinfo{volume}{93}}, \bibinfo{pages}{57001} (\bibinfo{year}{2011}).

\bibitem{Xu}
\bibinfo{author}{Xu, M.} \emph{et~al.}
\newblock \bibinfo{title}{{Evidence for an $s$-wave superconducting gap in
  {K}${}_{x}${F}e${}_{2\ensuremath{-}y}${S}e${}_{2}$ from angle-resolved
  photoemission}}.
\newblock \emph{\bibinfo{journal}{Phys. Rev. B}} \textbf{\bibinfo{volume}{85}},
  \bibinfo{pages}{220504} (\bibinfo{year}{2012}).

\bibitem{Wang_2012}
\bibinfo{author}{Wang, X.-P.} \emph{et~al.}
\newblock \bibinfo{title}{{Observation of an isotropic superconducting gap at
  the Brillouin zone centre of
  {T}l$_{0.63}${K}$_{0.37}${F}e$_{1.78}${S}e$_{2}$}}.
\newblock \emph{\bibinfo{journal}{Europhys. Lett.}}
  \textbf{\bibinfo{volume}{99}}, \bibinfo{pages}{67001} (\bibinfo{year}{2012}).

\bibitem{Pang}
\bibinfo{author}{Pang, G.} \emph{et~al.}
\newblock \bibinfo{title}{{Fully gapped d-wave superconductivity in
  {C}e{C}u$_2${S}i$_2$}}.
\newblock \emph{\bibinfo{journal}{Proc. Nat. Acad. Sci.}}
  \textbf{\bibinfo{volume}{115}}, \bibinfo{pages}{5343--5347}
  (\bibinfo{year}{2018}).

\bibitem{Kittaka}
\bibinfo{author}{Kittaka, S.} \emph{et~al.}
\newblock \bibinfo{title}{{Multiband superconductivity with unexpected
  deficiency of nodal quasiparticles in {C}e{C}u$_{2}${S}i$_{2}$}}.
\newblock \emph{\bibinfo{journal}{Phys. Rev. Lett.}}
  \textbf{\bibinfo{volume}{112}}, \bibinfo{pages}{067002}
  (\bibinfo{year}{2014}).

\bibitem{Yamashita}
\bibinfo{author}{Yamashita, T.} \emph{et~al.}
\newblock \bibinfo{title}{{Fully gapped superconductivity with no sign change
  in the prototypical heavy-fermion {C}e{C}u$_{2}${S}i$_{2}$}}.
\newblock \emph{\bibinfo{journal}{Sci. Adv.}} \textbf{\bibinfo{volume}{3}},
  \bibinfo{pages}{e1601667} (\bibinfo{year}{2017}).

\bibitem{Grinenko_Nat_Phys_2020}
\bibinfo{author}{Grinenko, V.} \emph{et~al.}
\newblock \bibinfo{title}{{Superconductivity with broken time-reversal symmetry
  inside a superconducting s-wave state}}.
\newblock \emph{\bibinfo{journal}{Nat. Phys.}} \textbf{\bibinfo{volume}{16}},
  \bibinfo{pages}{789--794} (\bibinfo{year}{2020}).

\bibitem{Zaki_arxiv_2019}
\bibinfo{author}{Zaki, N.}, \bibinfo{author}{Tsvelik, C., A.M.and~Wu} \&
  \bibinfo{author}{Johnson, P.}
\newblock \bibinfo{title}{Time reversal symmetry breaking in the
  Fe-chalcogenide superconductors. {P}reprint at
  \href{https://arxiv.org/abs/1907.11602}{https://arxiv.org/abs/1907.11602}}
  (\bibinfo{year}{2019}).

\bibitem{Goswami}
\bibinfo{author}{Goswami, P.}, \bibinfo{author}{Nikolic, P.} \&
  \bibinfo{author}{Si, Q.}
\newblock \bibinfo{title}{{Superconductivity in multi-orbital
  $t-{J}_{1}-{J}_{2}$ model and its implications for iron pnictides}}.
\newblock \emph{\bibinfo{journal}{Europhys. Lett.}}
  \textbf{\bibinfo{volume}{91}}, \bibinfo{pages}{37006} (\bibinfo{year}{2010}).

\bibitem{Wang_PRB_2010}
\bibinfo{author}{Wang, X.} \emph{et~al.}
\newblock \bibinfo{title}{Constraints imposed by symmetry on pairing operators
  for the iron pnictides}.
\newblock \emph{\bibinfo{journal}{Phys. Rev. B}} \textbf{\bibinfo{volume}{81}},
  \bibinfo{pages}{144509} (\bibinfo{year}{2010}).

\bibitem{Steg_1979}
\bibinfo{author}{Steglich, F.} \emph{et~al.}
\newblock \bibinfo{title}{{Superconductivity in the presence of strong Pauli
  paramagnetism: Ce${\mathrm{Cu}}_{2}{\mathrm{Si}}_{2}$}}.
\newblock \emph{\bibinfo{journal}{Phys. Rev. Lett.}}
  \textbf{\bibinfo{volume}{43}}, \bibinfo{pages}{1892} (\bibinfo{year}{1979}).

\bibitem{Mathur98}
\bibinfo{author}{Mathur, N.~D.} \emph{et~al.}
\newblock \bibinfo{title}{{Magnetically mediated superconductivity in heavy
  fermion compounds}}.
\newblock \emph{\bibinfo{journal}{Nature}} \textbf{\bibinfo{volume}{394}},
  \bibinfo{pages}{39--43} (\bibinfo{year}{1998}).

\bibitem{Gegenwart.08}
\bibinfo{author}{Gegenwart, P.}, \bibinfo{author}{Si, Q.} \&
  \bibinfo{author}{Steglich, F.}
\newblock \bibinfo{title}{{Quantum criticality in heavy-fermion metals}}.
\newblock \emph{\bibinfo{journal}{Nat.~Phys.}} \textbf{\bibinfo{volume}{4}},
  \bibinfo{pages}{186--197} (\bibinfo{year}{2008}).

\bibitem{Thompson.12}
\bibinfo{author}{Thompson, J.~D.} \& \bibinfo{author}{Fisk, Z.}
\newblock \bibinfo{title}{{Progress in heavy-fermion superconductivity: Ce115
  and related materials}}.
\newblock \emph{\bibinfo{journal}{J. Phys. Soc. Jpn.}}
  \textbf{\bibinfo{volume}{81}}, \bibinfo{pages}{011002}
  (\bibinfo{year}{2012}).

\bibitem{Ran684}
\bibinfo{author}{Ran, S.} \emph{et~al.}
\newblock \bibinfo{title}{{Nearly ferromagnetic spin-triplet
  superconductivity}}.
\newblock \emph{\bibinfo{journal}{Science}} \textbf{\bibinfo{volume}{365}},
  \bibinfo{pages}{684--687} (\bibinfo{year}{2019}).

\bibitem{Pustogow2019}
\bibinfo{author}{Pustogow, A.} \emph{et~al.}
\newblock \bibinfo{title}{{Constraints on the superconducting order parameter
  in Sr$_{2}$RuO$_{4}$ from oxygen-17 nuclear magnetic resonance}}.
\newblock \emph{\bibinfo{journal}{Nature}} \textbf{\bibinfo{volume}{574}},
  \bibinfo{pages}{72--75} (\bibinfo{year}{2019}).

\bibitem{Ishida1998}
\bibinfo{author}{Ishida, K.} \emph{et~al.}
\newblock \bibinfo{title}{{Spin-triplet superconductivity in Sr$_{2}$RuO$_{4}$
  identified by $^{17}$O Knight shift}}.
\newblock \emph{\bibinfo{journal}{Nature}} \textbf{\bibinfo{volume}{396}},
  \bibinfo{pages}{658--660} (\bibinfo{year}{2008}).

\bibitem{Mack_rmp2003}
\bibinfo{author}{Mackenzie, A.~P.} \& \bibinfo{author}{Maeno, Y.}
\newblock \bibinfo{title}{{The superconductivity of
  ${\mathrm{Sr}}_{2}{\mathrm{RuO}}_{4}$ and the physics of spin-triplet
  pairing}}.
\newblock \emph{\bibinfo{journal}{Rev. Mod. Phys.}}
  \textbf{\bibinfo{volume}{75}}, \bibinfo{pages}{657--712}
  (\bibinfo{year}{2003}).

\bibitem{Rice_1995}
\bibinfo{author}{Rice, T.~M.} \& \bibinfo{author}{Sigrist, M.}
\newblock \bibinfo{title}{{Sr$_{2}$RuO$_{4}$: an electronic analogue of
  $^{3}$He?}}
\newblock \emph{\bibinfo{journal}{J. Phys. Condens. Matter}}
  \textbf{\bibinfo{volume}{7}}, \bibinfo{pages}{L643--L648}
  (\bibinfo{year}{1995}).

\bibitem{Kallin_2009}
\bibinfo{author}{Kallin, C.} \& \bibinfo{author}{Berlinsky, A.~J.}
\newblock \bibinfo{title}{{Is Sr$_{2}$RuO$_{4}$ a chiral p-wave superconductor?}}
\newblock \emph{\bibinfo{journal}{J. Phys. Condens. Matter}}
  \textbf{\bibinfo{volume}{21}}, \bibinfo{pages}{164210}
  (\bibinfo{year}{2009}).

\bibitem{Ramires_ruthenate2019}
\bibinfo{author}{Ramires, A.} \& \bibinfo{author}{Sigrist, M.}
\newblock \bibinfo{title}{{Superconducting order parameter of
  ${\mathrm{Sr}}_{2}{\mathrm{RuO}}_{4}$: A microscopic perspective}}.
\newblock \emph{\bibinfo{journal}{Phys. Rev. B}}
  \textbf{\bibinfo{volume}{100}}, \bibinfo{pages}{104501}
  (\bibinfo{year}{2019}).

\bibitem{Huang2019}
\bibinfo{author}{Huang, W.}, \bibinfo{author}{Zhou, Y.} \&
  \bibinfo{author}{Yao, H.}
\newblock \bibinfo{title}{{Exotic Cooper pairing in multiorbital models of
  ${\mathrm{Sr}}_{2}{\mathrm{RuO}}_{4}$}}.
\newblock \emph{\bibinfo{journal}{Phys. Rev. B}}
  \textbf{\bibinfo{volume}{100}}, \bibinfo{pages}{134506}
  (\bibinfo{year}{2019}).

\bibitem{Raghu}
\bibinfo{author}{Raghu, S.}, \bibinfo{author}{Qi, X.-L.}, \bibinfo{author}{Liu,
  C.-X.}, \bibinfo{author}{Scalapino, D.~J.} \& \bibinfo{author}{Zhang, S.-C.}
\newblock \bibinfo{title}{{Minimal two-band model of the superconducting iron
  oxypnictides}}.
\newblock \emph{\bibinfo{journal}{Phys. Rev. B}} \textbf{\bibinfo{volume}{77}},
  \bibinfo{pages}{220503} (\bibinfo{year}{2008}).

\bibitem{Si_Abrahams2008}
\bibinfo{author}{Si, Q.} \& \bibinfo{author}{Abrahams, E.}
\newblock \bibinfo{title}{{Strong correlations and magnetic frustration in the
  high ${T}_{c}$ iron pnictides}}.
\newblock \emph{\bibinfo{journal}{Phys. Rev. Lett.}}
  \textbf{\bibinfo{volume}{101}}, \bibinfo{pages}{076401}
  (\bibinfo{year}{2008}).

\bibitem{Daghofer2010}
\bibinfo{author}{Daghofer, M.}, \bibinfo{author}{Nicholson, A.},
  \bibinfo{author}{Moreo, A.} \& \bibinfo{author}{Dagotto, E.}
\newblock \bibinfo{title}{Three orbital model for the iron-based
  superconductors}.
\newblock \emph{\bibinfo{journal}{Phys. Rev. B}} \textbf{\bibinfo{volume}{81}},
  \bibinfo{pages}{014511} (\bibinfo{year}{2010}).

\bibitem{Zhou_PRB_2008}
\bibinfo{author}{Zhou, Y.}, \bibinfo{author}{Chen, W.-Q.} \&
  \bibinfo{author}{Zhang, F.-C.}
\newblock \bibinfo{title}{{Symmetry of superconducting states with two orbitals
  on a tetragonal lattice: application to
  ${\text{LaFeAsO}}_{1\ensuremath{-}x}{\text{F}}_{x}$}}.
\newblock \emph{\bibinfo{journal}{Phys. Rev. B}} \textbf{\bibinfo{volume}{78}},
  \bibinfo{pages}{064514} (\bibinfo{year}{2008}).

\bibitem{Nicholson_PRB_2012}
\bibinfo{author}{Nicholson, A.} \emph{et~al.}
\newblock \bibinfo{title}{{Pairing symmetries of a hole-doped extended
  two-orbital model for the pnictides}}.
\newblock \emph{\bibinfo{journal}{Phys. Rev. B}} \textbf{\bibinfo{volume}{85}},
  \bibinfo{pages}{024532} (\bibinfo{year}{2012}).

\bibitem{Lv_PRB_2013}
\bibinfo{author}{Lv, W.}, \bibinfo{author}{Moreo, A.} \&
  \bibinfo{author}{Dagotto, E.}
\newblock \bibinfo{title}{${B}_{1g}$-like pairing states in two-leg ladder iron
  superconductors}.
\newblock \emph{\bibinfo{journal}{Phys. Rev. B}} \textbf{\bibinfo{volume}{88}},
  \bibinfo{pages}{094508} (\bibinfo{year}{2013}).

\bibitem{Graser.2009}
\bibinfo{author}{Graser, S.}, \bibinfo{author}{Maier, T.~A.},
  \bibinfo{author}{Hirschfeld, P.~J.} \& \bibinfo{author}{Scalapino, D.~J.}
\newblock \bibinfo{title}{{Near-degeneracy of several pairing channels in
  multiorbital models for the {F}e pnictides}}.
\newblock \emph{\bibinfo{journal}{New J. Phys.}} \textbf{\bibinfo{volume}{11}},
  \bibinfo{pages}{025016} (\bibinfo{year}{2009}).

\bibitem{Chubukov_PRB_2016}
\bibinfo{author}{Chubukov, A.~V.}, \bibinfo{author}{Vafek, O.} \&
  \bibinfo{author}{Fernandes, R.~M.}
\newblock \bibinfo{title}{Displacement and annihilation of Dirac gap nodes in
  $d$-wave iron-based superconductors}.
\newblock \emph{\bibinfo{journal}{Phys. Rev. B}} \textbf{\bibinfo{volume}{94}},
  \bibinfo{pages}{174518} (\bibinfo{year}{2016}).

\bibitem{Agterberg_PRL_2017}
\bibinfo{author}{Agterberg, D.~F.}, \bibinfo{author}{Shishidou, T.},
  \bibinfo{author}{O'Halloran, J.}, \bibinfo{author}{Brydon, P. M.~R.} \&
  \bibinfo{author}{Weinert, M.}
\newblock \bibinfo{title}{{Resilient nodeless $d$-wave superconductivity in
  monolayer {F}e{S}e}}.
\newblock \emph{\bibinfo{journal}{Phys. Rev. Lett.}}
  \textbf{\bibinfo{volume}{119}}, \bibinfo{pages}{267001}
  (\bibinfo{year}{2017}).

\bibitem{Leggett}
\bibinfo{author}{Leggett, A.~J.}
\newblock \bibinfo{title}{A theoretical description of the new phases of liquid
  $^{3}\mathrm{He}$}.
\newblock \emph{\bibinfo{journal}{Rev. Mod. Phys.}}
  \textbf{\bibinfo{volume}{47}}, \bibinfo{pages}{331--414}
  (\bibinfo{year}{1975}).

\bibitem{Vollhardt}
\bibinfo{author}{Vollhardt, D.} \& \bibinfo{author}{W\"{o}lfle, P.}
\newblock \emph{\bibinfo{title}{The superfluid phases of Helium 3}}
  (\bibinfo{publisher}{Taylor \& Francis, London}, \bibinfo{year}{1990}).

\bibitem{Yu}
\bibinfo{author}{Yu, R.}, \bibinfo{author}{Goswami, P.}, \bibinfo{author}{Si,
  Q.}, \bibinfo{author}{Nikolic, P.} \& \bibinfo{author}{Zhu, J.-X.}
\newblock \bibinfo{title}{{Superconductivity at the border of electron
  localization and itinerancy}}.
\newblock \emph{\bibinfo{journal}{Nat. Commun.}} \textbf{\bibinfo{volume}{4}},
  \bibinfo{pages}{2783} (\bibinfo{year}{2013}).

\bibitem{Sachdev}
\bibinfo{author}{Sachdev, S.}
\newblock \bibinfo{title}{{Quantum phase transitions of correlated electrons in
  two dimensions}}.
\newblock \emph{\bibinfo{journal}{Phys. A}} \textbf{\bibinfo{volume}{313}},
  \bibinfo{pages}{252 -- 283} (\bibinfo{year}{2002}).

\bibitem{Vieyra}
\bibinfo{author}{Vieyra, H.~A.} \emph{et~al.}
\newblock \bibinfo{title}{{Determination of gap symmetry from angle-dependent
  ${H}_{\text{c}2}$ measurements on {C}e{C}u$_{2}${S}i$_{2}$}}.
\newblock \emph{\bibinfo{journal}{Phys. Rev. Lett.}}
  \textbf{\bibinfo{volume}{106}}, \bibinfo{pages}{207001}
  (\bibinfo{year}{2011}).

\bibitem{Goremychkin}
\bibinfo{author}{Goremychkin, E.~A.} \& \bibinfo{author}{Osborn, R.}
\newblock \bibinfo{title}{Crystal-field excitations in
  {C}e{C}u$_{2}${S}i$_{2}$}.
\newblock \emph{\bibinfo{journal}{Phys. Rev. B}} \textbf{\bibinfo{volume}{47}},
  \bibinfo{pages}{14280} (\bibinfo{year}{1993}).

\bibitem{Rueff}
\bibinfo{author}{Rueff, J.-P.} \emph{et~al.}
\newblock \bibinfo{title}{{Absence of orbital rotation in superconducting
  {C}e{C}u$_{2}${G}e$_{2}$}}.
\newblock \emph{\bibinfo{journal}{Phys. Rev. B}} \textbf{\bibinfo{volume}{91}},
  \bibinfo{pages}{201108} (\bibinfo{year}{2015}).

\bibitem{Pourovskii}
\bibinfo{author}{Pourovskii, L.~V.}, \bibinfo{author}{Hansmann, P.},
  \bibinfo{author}{Ferrero, M.} \& \bibinfo{author}{Georges, A.}
\newblock \bibinfo{title}{{Theoretical prediction and spectroscopic
  fingerprints of an orbital transition in {C}e{C}u$_{2}${S}i$_{2}$}}.
\newblock \emph{\bibinfo{journal}{Phys. Rev. Lett.}}
  \textbf{\bibinfo{volume}{112}}, \bibinfo{pages}{106407}
  (\bibinfo{year}{2014}).

\bibitem{Koster}
\bibinfo{author}{Koster, G.~F.}
\newblock \emph{\bibinfo{title}{{Properties of the thirty-two point groups}}}
  (\bibinfo{publisher}{Cambridge, Mass., M.I.T. Press}, \bibinfo{year}{1963}).

\bibitem{Spille}
\bibinfo{author}{Spille, H.}, \bibinfo{author}{Rauchschwalbe, U.} \&
  \bibinfo{author}{Steglich, F.}
\newblock \bibinfo{title}{{Superconductivity in CeCu$_{2}$Si$_{2}$ : dependence
  of $T_{\text{c}}$ on alloying and stoichiometry}}.
\newblock \emph{\bibinfo{journal}{Helv. Phys. Acta}}
  \textbf{\bibinfo{volume}{56}}, \bibinfo{pages}{165--177}
  (\bibinfo{year}{1983}).

\bibitem{Yuan}
\bibinfo{author}{Yuan, H.~Q.} \emph{et~al.}
\newblock \bibinfo{title}{{Observation of two distinct superconducting phases
  in CeCu$_{2}$Si$_{2}$}}.
\newblock \emph{\bibinfo{journal}{Science}} \textbf{\bibinfo{volume}{302}},
  \bibinfo{pages}{2104--2107} (\bibinfo{year}{2003}).

\bibitem{Sigrist}
\bibinfo{author}{Sigrist, M.} \& \bibinfo{author}{Ueda, K.}
\newblock \bibinfo{title}{{Phenomenological theory of unconventional
  superconductivity}}.
\newblock \emph{\bibinfo{journal}{Rev. Mod. Phys.}}
  \textbf{\bibinfo{volume}{63}}, \bibinfo{pages}{239--311}
  (\bibinfo{year}{1991}).

\bibitem{Amorese_arxiv_2020}
\bibinfo{author}{Amorese, A.} \emph{et~al.}
\newblock \bibinfo{title}{Possible multi-orbital ground state in
  {C}e{C}u$_{2}${S}i$_{2}$. {P}reprint at
  \href{https://arxiv.org/abs/2010.01836}{https://arxiv.org/abs/2010.01836}}
  (\bibinfo{year}{2020}).

\end{thebibliography}

\begin{thebibliography}{10}
\expandafter\ifx\csname url\endcsname\relax
  \def\url#1{\texttt{#1}}\fi
\expandafter\ifx\csname urlprefix\endcsname\relax\def\urlprefix{URL }\fi
\providecommand{\bibinfo}[2]{#2}
\providecommand{\eprint}[2][]{\url{#2}}

\bibitem{de_Gennes}
\bibinfo{author}{de~Gennes, P.~G.}
\newblock \emph{\bibinfo{title}{Superconductivity of metals and alloys}}
  (\bibinfo{publisher}{Westview, Boulder}, \bibinfo{year}{1999}).

\bibitem{Vollhardt}
\bibinfo{author}{Vollhardt, D.} \& \bibinfo{author}{W\"{o}lfle, P.}
\newblock \emph{\bibinfo{title}{The superfluid phases of Helium 3}}
  (\bibinfo{publisher}{Taylor \& Francis, London}, \bibinfo{year}{1990}).

\bibitem{Anderson}
\bibinfo{author}{{Anderson, P.W. and Brinkman, W. F.}}
\newblock \bibinfo{title}{The theory of anisotropic superfluidity in
  {H}e$^{3}$ in \emph{Basic notions of condensed matter physics}} (\bibinfo{publisher}{Westview Press, U.S.},
  \bibinfo{year}{1997}).
  

\bibitem{Levin}
\bibinfo{author}{Levin, K.} \& \bibinfo{author}{Valls, O.}
\newblock \bibinfo{title}{{{Phenomenological theories of liquid $^{3}${H}e}}}.
\newblock \emph{\bibinfo{journal}{Phys. Rep.}} \textbf{\bibinfo{volume}{98}},
  \bibinfo{pages}{1 -- 56} (\bibinfo{year}{1983}).

\bibitem{Sigrist}
\bibinfo{author}{Sigrist, M.} \& \bibinfo{author}{Ueda, K.}
\newblock \bibinfo{title}{{Phenomenological theory of unconventional
  superconductivity}}.
\newblock \emph{\bibinfo{journal}{Rev. Mod. Phys.}}
  \textbf{\bibinfo{volume}{63}}, \bibinfo{pages}{239--311}
  (\bibinfo{year}{1991}).

\bibitem{Leggett}
\bibinfo{author}{Leggett, A.~J.}
\newblock \bibinfo{title}{A theoretical description of the new phases of liquid
  $^{3}\mathrm{He}$}.
\newblock \emph{\bibinfo{journal}{Rev. Mod. Phys.}}
  \textbf{\bibinfo{volume}{47}}, \bibinfo{pages}{331--414}
  (\bibinfo{year}{1975}).

\bibitem{Balian}
\bibinfo{author}{Balian, R.} \& \bibinfo{author}{Werthamer, N.~R.}
\newblock \bibinfo{title}{{Superconductivity with pairs in a relative $p$
  wave}}.
\newblock \emph{\bibinfo{journal}{Phys. Rev.}} \textbf{\bibinfo{volume}{131}},
  \bibinfo{pages}{1553--1564} (\bibinfo{year}{1963}).

\bibitem{Balatsky}
\bibinfo{author}{Balatskii, A.~V.}, \bibinfo{author}{Volovik, G.~E.} \&
  \bibinfo{author}{Konyshev, V.~A.}
\newblock \bibinfo{title}{{On the chiral anomaly in superfluid $^{3}$He-A}}.
\newblock \emph{\bibinfo{journal}{Zh. Eksp. Teor. Fiz.}}
  \textbf{\bibinfo{volume}{90}}, \bibinfo{pages}{2038--2056}
  (\bibinfo{year}{1986}).

\bibitem{Goswami}
\bibinfo{author}{Goswami, P.}, \bibinfo{author}{Nikolic, P.} \&
  \bibinfo{author}{Si, Q.}
\newblock \bibinfo{title}{{Superconductivity in multi-orbital
  $t-{J}_{1}-{J}_{2}$ model and its implications for iron pnictides}}.
\newblock \emph{\bibinfo{journal}{Europhys. Lett.}}
  \textbf{\bibinfo{volume}{91}}, \bibinfo{pages}{37006} (\bibinfo{year}{2010}).

\bibitem{Yu}
\bibinfo{author}{Yu, R.}, \bibinfo{author}{Goswami, P.}, \bibinfo{author}{Si,
  Q.}, \bibinfo{author}{Nikolic, P.} \& \bibinfo{author}{Zhu, J.-X.}
\newblock \bibinfo{title}{{Superconductivity at the border of electron
  localization and itinerancy}}.
\newblock \emph{\bibinfo{journal}{Nat. Commun.}} \textbf{\bibinfo{volume}{4}},
  \bibinfo{pages}{2783} (\bibinfo{year}{2013}).

\bibitem{Nica_Yu}
\bibinfo{author}{Nica, E.~M.}, \bibinfo{author}{Yu, R.} \& \bibinfo{author}{Si,
  Q.}
\newblock \bibinfo{title}{Orbital-selective pairing and superconductivity in
  iron selenides}.
\newblock \emph{\bibinfo{journal}{npj Quantum Materials}}
  \textbf{\bibinfo{volume}{2}}, \bibinfo{pages}{24} (\bibinfo{year}{2017}).
\newblock \eprint{arXiv:1505.04170}.

\bibitem{Lee_Zhang_Wu}
\bibinfo{author}{Lee, W.-C.}, \bibinfo{author}{Zhang, S.-C.} \&
  \bibinfo{author}{Wu, C.}
\newblock \bibinfo{title}{{Pairing state with a time-reversal symmetry breaking
  in {F}e{A}s-based superconductors}}.
\newblock \emph{\bibinfo{journal}{Phys. Rev. Lett.}}
  \textbf{\bibinfo{volume}{102}}, \bibinfo{pages}{217002}
  (\bibinfo{year}{2009}).

\bibitem{Raghu}
\bibinfo{author}{Raghu, S.}, \bibinfo{author}{Qi, X.-L.}, \bibinfo{author}{Liu,
  C.-X.}, \bibinfo{author}{Scalapino, D.~J.} \& \bibinfo{author}{Zhang, S.-C.}
\newblock \bibinfo{title}{{Minimal two-band model of the superconducting iron
  oxypnictides}}.
\newblock \emph{\bibinfo{journal}{Phys. Rev. B}} \textbf{\bibinfo{volume}{77}},
  \bibinfo{pages}{220503} (\bibinfo{year}{2008}).

\bibitem{Wan_Wang}
\bibinfo{author}{{Wan, Yuan and Wang, Qiang-Hua}}.
\newblock \bibinfo{title}{{Pairing symmetry and properties of iron-based
  high-temperature superconductors}}.
\newblock \emph{\bibinfo{journal}{Europhys. Lett.}}
  \textbf{\bibinfo{volume}{85}}, \bibinfo{pages}{57007} (\bibinfo{year}{2009}).

\bibitem{Schrieffer}
\bibinfo{author}{Schrieffer, J.~R.}
\newblock \emph{\bibinfo{title}{{Theory of Superconductivity}}}
  (\bibinfo{publisher}{Westview}, \bibinfo{address}{U.S.},
  \bibinfo{year}{1999}).

\end{thebibliography}
